\documentclass[twocolumn,preprint]{aastex62}

\usepackage{amsmath}
\graphicspath{{./}{figures/}}

\usepackage{color}

\usepackage{ulem}

\received{\today}
\revised{\today}
\accepted{\today}
\submitjournal{ApJ}

\shorttitle{All-Sky Anisotropy of Cosmic Rays at 10 TeV}
\shortauthors{HAWC Collaboration and IceCube Collaboration}

\begin{document}

\title{All-Sky Measurement of the Anisotropy of Cosmic Rays at 10 TeV \\
and Mapping of the Local Interstellar Magnetic Field}

\collaboration{HAWC Collaboration}
\thanks{Email: juan.diazvelez@alumnos.udg.mx}
\author{A.U.~Abeysekara}
\affiliation{Department of Physics and Astronomy, University of Utah, Salt Lake City, UT, USA}
\author{R.~Alfaro}
\affiliation{Instituto de F\'{i}sica, Universidad Nacional Aut\'{o}noma de M\'{e}xico, Ciudad de Mexico, Mexico}
\author{C.~Alvarez}
\affiliation{Universidad Aut\'{o}noma de Chiapas, Tuxtla Guti\'{e}rrez, Chiapas, M\'{e}xico}
\author{R.~Arceo}
\affiliation{Universidad Aut\'{o}noma de Chiapas, Tuxtla Guti\'{e}rrez, Chiapas, M\'{e}xico}
\author{J.C.~Arteaga-Vel\'{a}zquez}
\affiliation{Universidad Michoacana de San Nicol\'{a}s de Hidalgo, Morelia, Mexico}
\author{D.~Avila Rojas}
\affiliation{Instituto de F\'{i}sica, Universidad Nacional Aut\'{o}noma de M\'{e}xico, Ciudad de Mexico, Mexico}
\author{E.~Belmont-Moreno}
\affiliation{Instituto de F\'{i}sica, Universidad Nacional Aut\'{o}noma de M\'{e}xico, Ciudad de Mexico, Mexico}
\author{S.Y.~BenZvi}
\affiliation{Department~of Physics and Astronomy, University of Rochester, Rochester, NY 14627, USA}
\author{C.~Brisbois}
\affiliation{Department of Physics, Michigan Technological University, Houghton, MI, USA}
\author{T.~Capistr\'{a}n}
\affiliation{Instituto Nacional de Astrof\'{i}sica, \'Optica y Electr\'{o}nica, Puebla, Mexico}
\author{A.~Carramiñana}
\affiliation{Instituto Nacional de Astrof\'{i}sica, \'Optica y Electr\'{o}nica, Puebla, Mexico}
\author{S.~Casanova}
\affiliation{Institute of Nuclear Physics Polish Academy of Sciences, PL-31342 IFJ-PAN, Krakow, Poland}
\author{U.~Cotti}
\affiliation{Universidad Michoacana de San Nicol\'{a}s de Hidalgo, Morelia, Mexico}
\author{J.~Cotzomi}
\affiliation{Facultad de Ciencias F\'{i}sico Matem\'{a}ticas, Benem\'{e}rita Universidad Aut\'{o}noma de Puebla, Puebla, Mexico}
\author{J.C.~D\'{i}az-V\'{e}lez}
\affiliation{Departamento de F\'{i}sica, Centro Universitario de los Valles, Universidad de Guadalajara, Guadalajara, Mexico}
\affiliation{Department~of Physics and Wisconsin IceCube Particle Astrophysics Center, University of Wisconsin, Madison, WI 53706, USA}
\author{C.~De Le\'{o}n}
\affiliation{Facultad de Ciencias F\'{i}sico Matem\'{a}ticas, Benem\'{e}rita Universidad Aut\'{o}noma de Puebla, Puebla, Mexico}
\author{E.~De la Fuente}
\affiliation{Departamento de F\'{i}sica, Centro Universitario de Ciencias Exactase Ingenierias, Universidad de Guadalajara, Guadalajara, Mexico}
\author{S.~Dichiara}
\affiliation{Instituto de Astronom\'{i}a, Universidad Nacional Aut\'{o}noma de M\'{e}xico, Ciudad de Mexico, Mexico}
\author{M.A.~DuVernois}
\affiliation{Department~of Physics and Wisconsin IceCube Particle Astrophysics Center, University of Wisconsin, Madison, WI 53706, USA}
\author{C.~Espinoza}
\affiliation{Instituto de F\'{i}sica, Universidad Nacional Aut\'{o}noma de M\'{e}xico, Ciudad de Mexico, Mexico}
\author{D.W.~Fiorino}
\affiliation{Department~of Physics, University of Maryland, College Park, MD 20742, USA}
\author{H.~Fleischhack}
\affiliation{Department of Physics, Michigan Technological University, Houghton, MI, USA}
\author{N.~Fraija}
\affiliation{Instituto de Astronom\'{i}a, Universidad Nacional Aut\'{o}noma de M\'{e}xico, Ciudad de Mexico, Mexico}
\author{A.~Galv\'{a}n-G\'{a}mez}
\affiliation{Instituto de Astronom\'{i}a, Universidad Nacional Aut\'{o}noma de M\'{e}xico, Ciudad de Mexico, Mexico}
\author{J.A.~Garc\'{i}a-Gonz\'{a}lez}
\affiliation{Instituto de F\'{i}sica, Universidad Nacional Aut\'{o}noma de M\'{e}xico, Ciudad de Mexico, Mexico}
\author{M.M.~Gonz\'{a}lez}
\affiliation{Instituto de Astronom\'{i}a, Universidad Nacional Aut\'{o}noma de M\'{e}xico, Ciudad de Mexico, Mexico}
\author{J.A.~Goodman}
\affiliation{Department~of Physics, University of Maryland, College Park, MD 20742, USA}
\author{Z.~Hampel-Arias}
\affiliation{Department~of Physics and Wisconsin IceCube Particle Astrophysics Center, University of Wisconsin, Madison, WI 53706, USA}
\affiliation{Inter-university Institute for High Energies, Universit\'{e} Libre de Bruxelles, Bruxelles, Belgium}
\author{J.P.~Harding}
\affiliation{Physics Division, Los Alamos National Laboratory, Los Alamos, NM, USA}
\author{S.~Hernandez}
\affiliation{Instituto de F\'{i}sica, Universidad Nacional Aut\'{o}noma de M\'{e}xico, Ciudad de Mexico, Mexico}
\author{B.~Hona}
\affiliation{Department of Physics, Michigan Technological University, Houghton, MI, USA}
\author{F.~Hueyotl-Zahuantitla}
\affiliation{Universidad Aut\'{o}noma de Chiapas, Tuxtla Guti\'{e}rrez, Chiapas, M\'{e}xico}
\author{A.~Iriarte}
\affiliation{Instituto de Astronom\'{i}a, Universidad Nacional Aut\'{o}noma de M\'{e}xico, Ciudad de Mexico, Mexico}
\author{A.~Jardin-Blicq}
\affiliation{Max-Planck Institute for Nuclear Physics, 69117 Heidelberg, Germany}
\author{V.~Joshi}
\affiliation{Max-Planck Institute for Nuclear Physics, 69117 Heidelberg, Germany}
\author{A.~Lara}
\affiliation{Instituto de Geof\'{i}sica, Universidad Nacional Aut\'{o}noma de M\'{e}xico, Ciudad de Mexico, Mexico}
\author{H.~Le\'{o}n Vargas}
\affiliation{Instituto de F\'{i}sica, Universidad Nacional Aut\'{o}noma de M\'{e}xico, Ciudad de Mexico, Mexico}
\author{G.~Luis-Raya}
\affiliation{Universidad Politecnica de Pachuca, Pachuca, Hgo, Mexico}
\author{K.~Malone}
\affiliation{Department~of Physics, Pennsylvania State University, University Park, PA 16802, USA}
\author{S.S.~Marinelli}
\affiliation{Department~of Physics and Astronomy, Michigan State University, East Lansing, MI 48824, USA}
\author{J.~Mart\'{i}nez-Castro}
\affiliation{Centro de Investigaci\'on en Computaci\'on, Instituto Polit\'{e}cnico Nacional, M\'{e}xico City, M\'{e}xico.}
\author{O.~Martinez}
\affiliation{Facultad de Ciencias F\'{i}sico Matem\'{a}ticas, Benem\'{e}rita Universidad Aut\'{o}noma de Puebla, Puebla, Mexico}
\author{J.A.~Matthews}
\affiliation{Departmentof Physics and Astronomy, University of New Mexico, Albuquerque, NM, USA}
\author{P.~Miranda-Romagnoli}
\affiliation{Universidad Aut\'{o}noma del Estado de Hidalgo, Pachuca, Mexico}
\author{E.~Moreno}
\affiliation{Facultad de Ciencias F\'{i}sico Matem\'{a}ticas, Benem\'{e}rita Universidad Aut\'{o}noma de Puebla, Puebla, Mexico}
\author{M.~Mostaf\'{a}}
\affiliation{Department~of Physics, Pennsylvania State University, University Park, PA 16802, USA}
\author{L.~Nellen}
\affiliation{Instituto de Ciencias Nucleares, Universidad Nacional Aut\'{o}noma de Mexico, Ciudad de Mexico, Mexico}
\author{M.~Newbold}
\affiliation{Department of Physics and Astronomy, University of Utah, Salt Lake City, UT, USA}
\author{M.U.~Nisa}
\affiliation{Department~of Physics and Astronomy, University of Rochester, Rochester, NY 14627, USA}
\author{R.~Noriega-Papaqui}
\affiliation{Universidad Aut\'{o}noma del Estado de Hidalgo, Pachuca, Mexico}
\author{E.G.~P\'{e}rez-P\'{e}rez}
\affiliation{Universidad Politecnica de Pachuca, Pachuca, Hgo, Mexico}
\author{J.~Pretz}
\affiliation{Department~of Physics, Pennsylvania State University, University Park, PA 16802, USA}
\author{Z.~Ren}
\affiliation{Departmentof Physics and Astronomy, University of New Mexico, Albuquerque, NM, USA}
\author{C.D.~Rho}
\affiliation{Department~of Physics and Astronomy, University of Rochester, Rochester, NY 14627, USA}
\author{C.~Rivi\`{e}re}
\affiliation{Department~of Physics, University of Maryland, College Park, MD 20742, USA}
\author{D.~Rosa-Gonz\'{a}lez}
\affiliation{Instituto Nacional de Astrof\'{i}sica, \'Optica y Electr\'{o}nica, Puebla, Mexico}
\author{M.~Rosenberg}
\affiliation{Department~of Physics, Pennsylvania State University, University Park, PA 16802, USA}
\author{H.~Salazar}
\affiliation{Facultad de Ciencias F\'{i}sico Matem\'{a}ticas, Benem\'{e}rita Universidad Aut\'{o}noma de Puebla, Puebla, Mexico}
\author{F.~Salesa Greus}
\affiliation{Institute of Nuclear Physics Polish Academy of Sciences, PL-31342 IFJ-PAN, Krakow, Poland}
\author{A.~Sandoval}
\affiliation{Instituto de F\'{i}sica, Universidad Nacional Aut\'{o}noma de M\'{e}xico, Ciudad de Mexico, Mexico}
\author{M.~Schneider}
\affiliation{Department~of Physics, University of Maryland, College Park, MD 20742, USA}
\author{H.~Schoorlemmer}
\affiliation{Max-Planck Institute for Nuclear Physics, 69117 Heidelberg, Germany}
\author{G.~Sinnis}
\affiliation{Physics Division, Los Alamos National Laboratory, Los Alamos, NM, USA}
\author{A.J.~Smith}
\affiliation{Department~of Physics, University of Maryland, College Park, MD 20742, USA}
\author{P.~Surajbali}
\affiliation{Max-Planck Institute for Nuclear Physics, 69117 Heidelberg, Germany}
\author{I.~Taboada}
\affiliation{School of Physics and Center for Relativistic Astrophysics, Georgia Institute of Technology, Atlanta, GA 30332, USA}
\author{K.~Tollefson}
\affiliation{Department~of Physics and Astronomy, Michigan State University, East Lansing, MI 48824, USA}
\author{I.~Torres}
\affiliation{Instituto Nacional de Astrof\'{i}sica, \'Optica y Electr\'{o}nica, Puebla, Mexico}
\author{L.~Villaseñor}
\affiliation{Facultad de Ciencias F\'{i}sico Matem\'{a}ticas, Benem\'{e}rita Universidad Aut\'{o}noma de Puebla, Puebla, Mexico}
\author{T.~Weisgarber}
\affiliation{Department~of Physics and Wisconsin IceCube Particle Astrophysics Center, University of Wisconsin, Madison, WI 53706, USA}
\author{J.~Wood}
\affiliation{Department~of Physics and Wisconsin IceCube Particle Astrophysics Center, University of Wisconsin, Madison, WI 53706, USA}
\author{A.~Zepeda}
\affiliation{Physics Department, Centro de Investigacion y de Estudios Avanzados del IPN, Mexico City, DF, Mexico}
\author{H.~Zhou}
\affiliation{Physics Division, Los Alamos National Laboratory, Los Alamos, NM, USA}
\author{J.D.~\'{A}lvarez}
\affiliation{Universidad Michoacana de San Nicol\'{a}s de Hidalgo, Morelia, Mexico}

\collaboration{IceCube Collaboration}
\thanks{Email: analysis@icecube.wisc.edu}

\author{M.~G.~Aartsen}
\affiliation{Department~of Physics and Astronomy, University of Canterbury, Private Bag 4800, Christchurch, New Zealand}
\author{M.~Ackermann}
\affiliation{DESY, D-15738 Zeuthen, Germany}
\author{J.~Adams}
\affiliation{Department~of Physics and Astronomy, University of Canterbury, Private Bag 4800, Christchurch, New Zealand}
\author{J.~A.~Aguilar}
\affiliation{Universit\'e Libre de Bruxelles, Science Faculty CP230, B-1050 Brussels, Belgium}
\author{M.~Ahlers}
\affiliation{Niels Bohr Institute, University of Copenhagen, DK-2100 Copenhagen, Denmark}
\author{M.~Ahrens}
\affiliation{Oskar Klein Centre and Department~of Physics, Stockholm University, SE-10691 Stockholm, Sweden}
\author{D.~Altmann}
\affiliation{Erlangen Centre for Astroparticle Physics, Friedrich-Alexander-Universit\"at Erlangen-N\"urnberg, D-91058 Erlangen, Germany}
\author{K.~Andeen}
\affiliation{Department of Physics, Marquette University, Milwaukee, WI, 53201, USA}
\author{T.~Anderson}
\affiliation{Department~of Physics, Pennsylvania State University, University Park, PA 16802, USA}
\author{I.~Ansseau}
\affiliation{Universit\'e Libre de Bruxelles, Science Faculty CP230, B-1050 Brussels, Belgium}
% ...




\begin{abstract}
We present  the first full-sky analysis of the cosmic ray arrival direction distribution with data collected by the High-Altitude Water Cherenkov and IceCube observatories in the northern and southern hemispheres at the same median primary particle energy of 10 TeV. 
The combined sky map and angular power spectrum largely eliminate biases that result from partial sky coverage and present a key to probe into the propagation properties of TeV cosmic rays through our local interstellar medium and the interaction between the interstellar and heliospheric magnetic fields.
From the map we determine the horizontal dipole components of the anisotropy
$\delta_{0h} =  9.16 \times 10^{-4}$ and
$\delta_{6h} = 7.25 \times 10^{-4}~(\pm0.04 \times 10^{-4})$. 
In addition, we infer the direction 
($229.2\pm 3^\circ.5$ R.A. , $11.4\pm 3^\circ.0$ decl.) 
of the interstellar magnetic field from the boundary between large scale excess and deficit regions from which we estimate the missing corresponding vertical dipole component of the large scale anisotropy
to be $\delta_N \sim -3.97 ^{+1.0}_{-2.0} \times 10^{-4}$.
\end{abstract}



\section{Introduction} \label{sec:intro}
A number of theoretical models predict an anisotropy in the distribution of arrival directions of cosmic rays that results from the distribution of sources in the Galaxy and diffusive propagation of these particles~\citep{erlykin_2006, Blasi:2012jan, Ptuskin:2012dec, Pohl:2013mar, 
sveshnikova_2013, Kumar:2014apr, Mertsch:2015jan}. 
Although the observed distribution of cosmic rays is highly isotropic, several ground-based experiments located either in the northern or southern hemisphere have observed small but significant variations in the arrival direction distribution of TeV to PeV cosmic rays with high statistical accuracy, in both large and medium angular scales~\citep{nagashima_1998,hall_1999,amenomori_2005,Amenomori:2006bx,guillian_2007,Milagro:2008nov,abdo_2009,aglietta_2009,munakata_2010,IceCube:2010aug,Abbasi:2011ai,MINOS:2011icrc,IceCube:2012feb,IceCube:2013mar,ARGO-YBJ:2013gya,HAWC:2014dec,bartoli_2015,icecube2016,tibet2017,bartoli2018,hawclsa2017}.
The observed large-scale anisotropy has an amplitude of about 
$10^{-3}$ and small-scale structures of amplitude of $10^{-4}$ with angular size of $10^\circ$ to $30^\circ$.

For previously reported measurements that rely on time-integrated methods~\citep{Alexandreas1993,Atkins:2003ep}, a difference between the instantaneous and integrated field of view of the experiments can lead to an attenuation of structures with angular size larger than the instantaneous field of view~\citep{0004-637X-823-1-10}. For this analysis, we apply an optimal reconstruction method that can recover the amplitude of the projected large-scale anisotropy.
The limited integrated field of view of the sky in all of these individual measurements also makes it difficult to 
correctly characterize such an anisotropy in terms of its spherical harmonic components and
produce a quantitative measurement of the large scale characteristics, such as its dipole or quadrupole component, without a high degree of degeneracy~\citep{SOMMERS2001271}. 
The resulting correlations between the multipole spherical harmonic terms $a_{\ell m}$ bias the interpretation of the cosmic ray distributions in the context of particle diffusion in the local interstellar medium (LISM). 
In this joint analysis by the High-Altitude Water Cherenkov
(HAWC) and IceCube collaborations we have 
combined data from both experiments at 10 TeV median primary particle energy
to study the full-sky anisotropy.
Important information can be obtained from the power spectrum of the spherical harmonic components at low $\ell$ (large scale), 
which is most affected by partial sky coverage.
It should be noted that neither observatory is sensitive to variations across decl. bands since events recorded from a fixed direction in the local coordinate system can only probe the cosmic-ray flux in a fixed decl. band $\delta$.
As a result, the dipole anisotropy can only be observed as a projection onto the equatorial plane. However, some information about the vertical component can be inferred from medium- and small-scale structures.

\section{The HAWC Detector}\label{sec:HAWC}
The HAWC Gamma-Ray Observatory is an
extensive air-shower detector array located at 4100\,m a.s.l. on the slopes
of Volc\'{a}n Sierra Negra at $19^{\circ}$N in the state of Puebla, Mexico.
While HAWC is designed to study the sky in gamma rays between 500\,GeV and 100\,TeV,  
it is also sensitive to showers from primary cosmic rays up to multi-PeV energies with an instantaneous field of view of about $2$\,sr. 

The detector consists of a 22,000 m$^2$ array of 300 close-packed water Cherenkov detectors (WCDs), 
each containing 200 metric tons of purified water and four upward-facing photomultiplier tubes (PMTs). 
At the bottom of each WCD, three 8-inch Hamamatsu R5912 
photomultiplier tubes (PMTs) are anchored in an equilateral
triangle of side length 3.2 meters, with one 10-inch high-quantum efficiency
Hamamatsu R7081 PMT anchored at the center.

As secondary air shower particles pass through the WCDs, the produced Cherenkov light is collected
by the PMTs, permitting the reconstruction of primary particle properties including the local arrival direction,
core location, and the energy. 
Further details on the HAWC detector can be found in~\cite{Abeysekara:2017mjj,ABEYSEKARA2018138}.

The light-tight nature of the WCDs allows the detector to operate at nearly 100\% up-time efficiency,
with the data acquisition system recording air showers at a rate of $\sim$ 25 kHz. 
With a resulting daily sky coverage of $8.4$\,sr and an angular resolution of $0^{\circ}.4$ for energies above 10 TeV, HAWC is an ideal instrument for measuring the cosmic-ray arrival direction distribution with unprecedented precision.

\section{The IceCube Detector}\label{sec:icecube}
The IceCube Neutrino Observatory, located at the 
geographic South Pole, is composed of a neutrino detector in the deep ice and a surface air-shower array.
The in-ice IceCube detector consists of 86
vertical strings containing a total of 5,160 optical sensors, called digital
optical modules (DOMs), frozen in the ice at depths from 
1,450 meters to 2,450 meters
below the surface of the ice.  A DOM consists of a pressure-protective glass 
sphere that houses a 10-inch Hamamatsu photomultiplier tube together with 
electronic boards used for detection, digitization, and readout.  The strings 
are separated by an average distance of 125\,m, each one hosting 60 DOMs equally 
spaced over the kilometer of instrumented length.  The DOMs detect Cherenkov 
radiation produced by relativistic particles passing through the ice, including
muons and muon bundles produced by cosmic-ray air showers in the 
atmosphere above IceCube.  These atmospheric muons form a large background for 
neutrino analyses, but also provide an opportunity to use IceCube as a 
large cosmic-ray detector. Further details on the IceCube detector can be found in~\cite{IceCube:detector}. 

All events that trigger IceCube are reconstructed using a likelihood-based method that
accounts for light propagation in the 
ice~\citep{Ahrens2004169}.  The fit provides a median angular resolution of $3^{\circ}$ according to simulation~\citep{Abbasi:2011ai} 
but worsens past zenith 
angles of approximately $70^{\circ}$~\cite{IceCube:detector}. 
This is not to be confused with the $\sim 0^{\circ}.6$ angular resolution of IceCube for neutrino-induced tracks of where more sophisticated reconstruction algorithms and more stringent quality cuts are applied. 
The energy threshold of cosmic-ray primaries producing atmospheric muons in IceCube is limited by the minimum muon energy required to penetrate the ice. As a result, the primary particle energy threshold increases with larger zenith angles as muons must travel increasingly longer distances through the ice. This is accounted for in the analysis as described in Sec. \ref{sec:dataset}.
Due to the limited data transfer rate available from the South Pole,
cosmic-ray induced muon data are stored in a compact data storage and transfer (DST) format~\citep{Abbasi:2011ai},
containing the results of the angular reconstructions described as well as
some limited information per event. However, detailed information such as PMT waveforms used for these reconstructions is not kept. The preliminary reconstructions encoded in the DST rely on faster, less accurate methods than those applied to the filtered dataset used in most neutrino analyses.

\section{The Dataset}\label{sec:dataset}
The dataset selected for this analysis is composed of 5 years of data collected by the IceCube Neutrino Observatory between May 2011 and May 2016, as well as 2 years of data from the HAWC Gamma-Ray Observatory collected between May 2015 and May 2017. 
In order to reduce bias from uneven exposure along R.A., only full sidereal days of continuous data-taking were chosen for this study. The residual contribution of the dipole anisotropy induced by the motion of the Earth around the Sun is estimated to be on the order of $10^{-5}$, which is smaller than the statistical error of this analysis (see section \ref{sec:systematics}).
Cuts are applied to each dataset in order to improve the angular resolution and energy resolution of reconstructed events.
In the case of HAWC these include a cut on the number of active optical sensors in order to increase the information available for the reconstruction of the shower. A cut on the reconstructed zenith angle excludes events with $\theta > 57^\circ$ where the quality of reconstructions decreases rapidly.  A cut is also applied on the variable CxPE40 which corresponds to the effective charge measured in the PMT with the largest effective charge at a distance of more than 40 m from the shower core with CxPE40 $> 40$. The effective charge $Q_\mathrm{eff}$ scales the charge of higher-efficiency central 10-inch PMTs by a factor of 0.46 relative to the 8-inch PMTs so that all optical sensors are treated equally. The value of CxPE40 is typically large for a hadronic events~\citep{Abeysekara:2017mjj}.
Finally, in order to identify and exclude gamma-ray candidates, a cut is applied on $\mathcal {P}$, that describes the ``clumpiness'' of the air shower \citep{Abeysekara:2017mjj} with $\mathcal {P}> 1.8$. 
$\mathcal{P}$ is defined using
the lateral distribution function of the air shower.
$\mathcal{P}$ is computed using the logarithm of the effective charge
$\zeta_{i}={\rm{log}}_{10}(Q_{\rm{eff},i})$. For each PMT hit, $i$, an expectation
is assigned $\langle\zeta_i\rangle$ 
by averaging the $\zeta_i$ in all PMTs
contained in an annulus containing the hit, with a width of 5 meters, 
centered at the core of the air shower. 
$\mathcal{P}$ is then calculated using the $\chi^2$ formula:
\begin{equation}\label{eq:pinc}
\mathcal{P}= {1 \over N} {\sum_{i=0}^{N}  { {(\zeta_i - \langle\zeta_i\rangle)^2} \over{ {\sigma_{\zeta_i}}^2}   }}
\end{equation}
\noindent The errors $\sigma_{\zeta_i}$ are assigned from a study of a sample strong gamma-ray candidates in the vicinity of the Crab nebula.
The $\mathcal{P}$ variable essentially requires axial smoothness.

In the case of IceCube we apply a cut on the reduced likelihood of the directional reconstruction ($\mathrm{RlogL} < 15$), defined as the best-fit log-likelihood divided by the number of degrees of freedom in the fit~\citep{Ahrens2004169} which gives an estimate of the goodness of fit for the angular reconstruction. There is also a cut on the number of \emph{direct} photoelectrons and the corresponding length of the track  $N_\mathrm{dir} > 9 \cos(\theta)$ and $l_\mathrm{dir}> 200 \cos(\theta)$~meters. This cut depends on the reconstructed zenith angle $\theta$ in order to preserve sufficient statistics near the horizon. 
Photons are considered direct when the time \emph{residual} (i.e., the delay in their arrival time due to scattering in the ice) falls within a time window of -15 ns to +75 ns with respect to the geometrically expected arrival time from the reconstructed track~\citep{Ahrens2004169}.

\begin{table*}[ht]
\centering  
\scriptsize
\begin{tabular}{p{0.19\textwidth}|p{0.10\textwidth} | p{0.13\textwidth} | p{0.10\textwidth} | p{0.13\textwidth}}
 &\multicolumn{2}{|l}{\small \bf{IceCube}} & \multicolumn{2}{|l}{\small \bf {HAWC}} \\  \hline
Latitude& \multicolumn{2}{|l}{90$^\circ$ S} &  \multicolumn{2}{|l}{19$^\circ $ N}\\
Detection method &  \multicolumn{2}{|l}{muons produced by CR}  &  \multicolumn{2}{|l}{air showers produced by CR and $\gamma$}  \\
Field of view  &  \multicolumn{2}{|l}{-90$^\circ$/-16$^\circ$ ($\delta$), ${\sim}$4 sr (same sky over 24h)} &  \multicolumn{2}{|l}{-30$^\circ$/68$^\circ$ ($\delta$),  ${\sim}$2 sr (8 sr observed/24 h)}\\
Livetime & \multicolumn{2}{|l}{1742 days over a period of 1826 days}  &  \multicolumn{2}{|l}{519 days over a period of 653 days} \\
Detector trigger rate & \multicolumn{2}{|l}{2.5 kHz} &  \multicolumn{2}{|l}{25 kHz}   \\ \hline
 & Quality cuts & Energy and quality cuts &  Quality cuts & Energy and quality cuts \\ \hline
Median primary energy & $20$ TeV & $10$ TeV &  $2$ TeV & $10$ TeV \\
Approx. angular resolution &  $2^\circ - 3^\circ$ & $2^\circ - 6^\circ$ &  $0.4^\circ - 0^\circ.8$ & $0.4^\circ - 1^\circ.0$ \\
Events & $2.8 \times 10^{11}$ & $1.7 \times 10^{11}$ & $7.1  \times 10^{10}$  & $2.8 \times 10^{10}$
\end{tabular}
\caption{
\small Comparison of the IceCube and HAWC datasets. The median primary particle energy, angular resolution and number of remaining events is shown in the sub-columns after applying only quality cuts and after applying both energy and quality cuts. The angular resolution of IceCube corresponds to the DST dataset that relies on faster, less accurate reconstructions as well as less stringent quality cuts. The energy cuts applied are chosen to lower the median energy of IceCube data from 20 TeV down to 10 TeV. In the case of HAWC, the cuts are aimed to raise the median energy of HAWC data from 2 TeV up to 10 TeV.
} \label{tab:observatories}
\end{table*}

Table \ref{tab:observatories} shows the characteristics of both experiments next to each other. 
The two detectors have different energy responses and this results in a difference in the median energy.
In order to select data that are consistent between the two detectors, we have applied additional cuts on the reconstructed energy of events: 
in the case of HAWC we use an energy reconstruction based on the likelihood method described in~\cite{hamphel2017} to select 
events with reconstructed energies at or above 10 TeV. 
In the case of IceCube we apply a cut in the two dimensional plane of number of hit optical sensors (which act as a proxy for muon energy) and the cosine of the reconstructed zenith angle, as described in \cite{IceCube:2012feb}. 
As a result of the overburden of ice described in Sec.~\ref{sec:icecube}, for a given number of hit optical sensors, events at larger zenith angles are produced by cosmic-ray particles with higher energy~\citep{IceCube:2012feb,icecube2016}.
The energy resolution is primarily limited by the relatively
large fluctuations in the fraction of the total shower energy
that is transferred to the muon bundle and is of the order of 0.5 in
$\log_{10}(E/\mathrm{GeV})$~\citep{icecube2016}. 

\begin{figure}[h]
\begin{center}
\includegraphics[width=0.98\columnwidth]{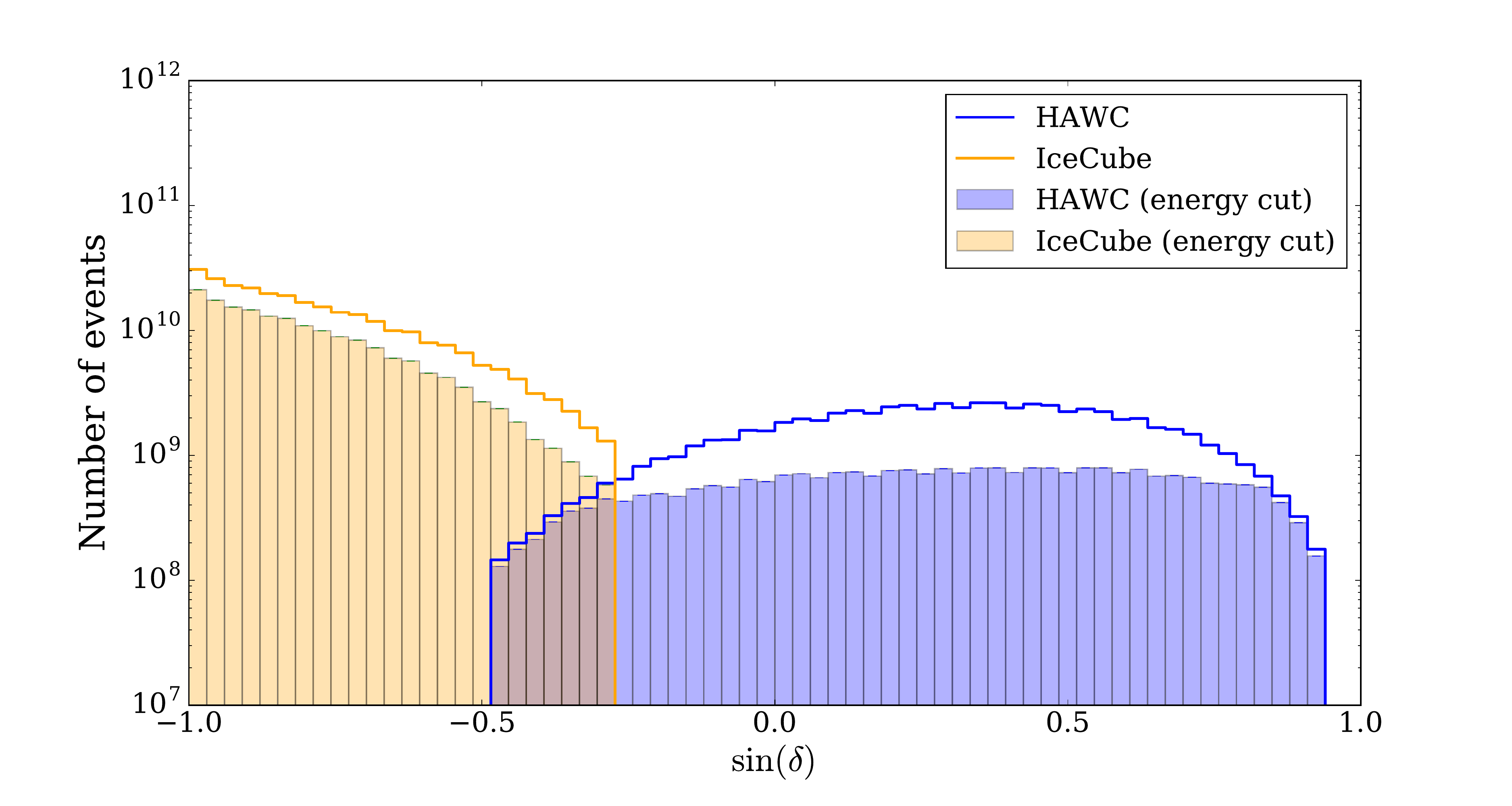}
    \caption{Distribution of events as a function of decl. for IceCube and HAWC. The figure shows the two datasets before and after applying energy and quality cuts. Restricting datasets to overlapping energy bins significantly reduces statistics for HAWC. The rates are dominated by events with energies near the threshold of each detector.
By imposing an artificial cut on low energies in the HAWC data, the detector response flattens since it becomes less dependent of zenith angle.
The statistics in HAWC with 300 tanks before cuts are comparable to one year of IceCube with 86 strings.}
        \label{fig:detector_acceptance}
        \end{center}
\end{figure}
\begin{figure}[h]
\begin{center}
\includegraphics[width=0.98\columnwidth]
{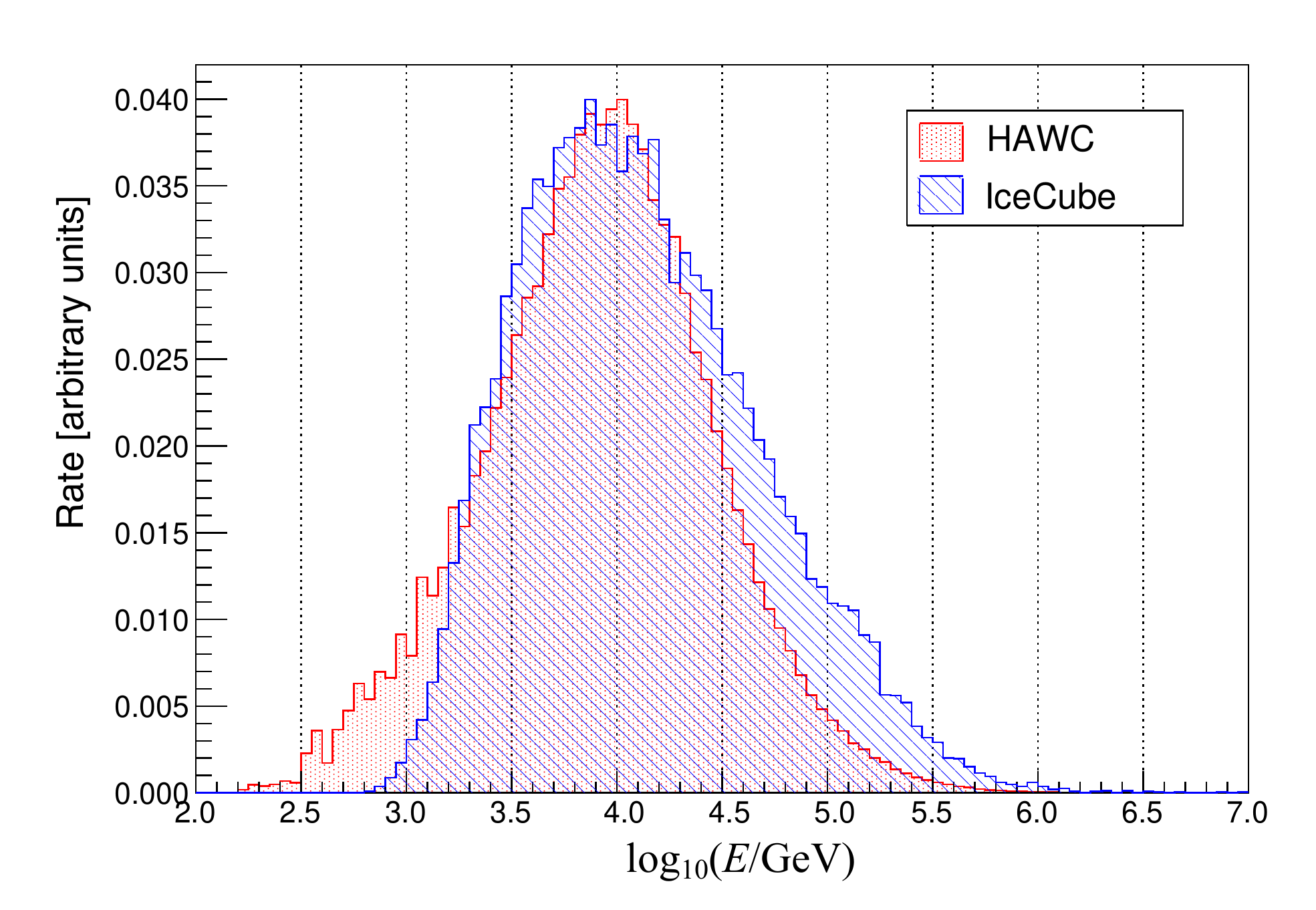}
              \caption{
	Energy distribution of the final event selection for the two datasets based on Monte Carlo simulations.            
              }
        \label{fig:energy_dist}
\end{center}
\end{figure}
\begin{figure}[h]
\begin{center}
\includegraphics[width=0.98\columnwidth]{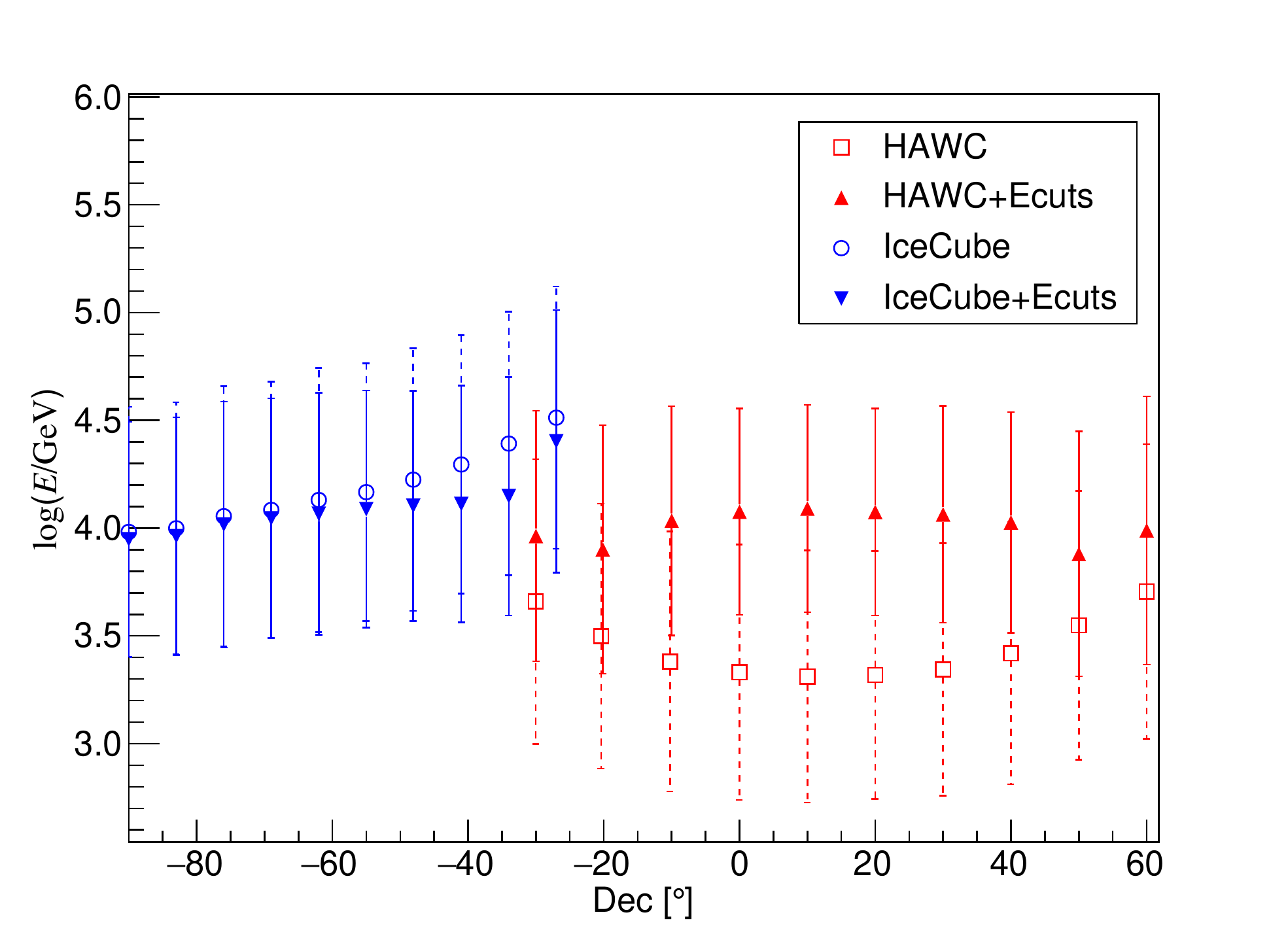}
\caption{Median energy as a function of decl. for Monte Carlo simulations before and after applying energy cuts.}
        \label{fig:energy_dec}
\end{center}
\end{figure}
Figure~\ref{fig:detector_acceptance} shows the distribution of data as a function of decl. 
The resulting energy distribution of the two datasets is shown in Figure~\ref{fig:energy_dist}. 
As a result of the applied energy cuts, both cosmic-ray data sets have a median primary particle energy of approximately 10 TeV with little dependence on zenith angle (Figure~\ref{fig:energy_dec}). The energy response of the observatories covers a 68\% range of approximately 3 TeV - 40 TeV, in the case of IceCube, and 2.5 TeV - 30 TeV for HAWC around the median energy.

The two experiments have different response to the cosmic ray mass composition. This is largely due to the detection method. 
Particles entering Earth's atmosphere (15 to 20 km above sea level) interact with nuclei in air and produce a cascade of secondary particles. This particle cascade continues to grow until ionization becomes the dominant energy loss mechanism. The depth X$_\mathrm{max}$ at which this happens depends on both the energy of the primary particle, and its mass. Lighter nuclei penetrate deeper than heavier nuclei. As a result, the altitude of extended air shower arrays such as HAWC 
can affect the response of the detector to different nuclei since they are sensitive to the electromagnetic component of the particle shower. In contrast, the IceCube in-ice detector observes cosmic rays through the detection of deep penetrating muons produced from the decay of charged pions and kaons generated in the early interactions. 
As a a result, for the same composition, IceCube's response to different cosmic-ray nuclei differs from that of HAWC. 

If the first interaction occurs at a lower air density (and higher elevation), mesons are more likely to decay to muons (and neutrinos) instead of re-interacting and producing lower energy pions and other secondary particles. As a result, the two experiments react differently to changes in atmospheric temperature and pressure.
\begin{table}[h]
\centering  
\begin{tabular}{l|l|l}
 & IceCube (10 TeV) & HAWC (10 TeV)  \\ \hline
Proton & 0.756 $\pm$ 0.018 & 0.6160 $\pm$ 0.0054 \\
He & 0.195 $\pm$ 0.009 & 0.3110 $\pm$ 0.0014 \\
CNO & 0.028 $\pm$ 0.004 & 0.0467 $\pm$ 0.0004 \\
NeMgSi & 0.013 $\pm$ 0.002 & 0.0191 $\pm$ 0.0001  \\
Fe & 0.008 $\pm$ 0.002 & 0.0078 $\pm$ 0.0001 \\
\end{tabular}
\caption{
\small Relative mass composition for 10 TeV median energy cosmic-rays in the two samples as determined from {\tt CORSIKA} Monte Carlo simulations~\citep{corsika} 
weighted to a Polygonato spectrum~\citep{horandel}. 
Errors reflect statistical uncertainties in the simulation datasets.
} \label{tab:composition}
\end{table} 
The data from both experiments are dominated by light nuclei (protons and alpha particles) as can be seen in Table \ref{tab:composition}. 
All of the cuts applied were chosen based on {\tt CORSIKA} Monte Carlo simulations~\citep{corsika} weighted to a Polygonato spectrum~\citep{horandel} and detailed simulations of the detector response.

\section{Analysis}
We compute the relative intensity as a function of J2000 equatorial
coordinates ($\alpha$, $\delta$) by binning the sky into an
equal-area grid with a bin size of $0^\circ.9$ using the 
{\tt HEALPix} library~\citep{Gorski:2005apr}. 
The angular distribution can be expressed as $\phi(\alpha, \delta) = \phi^\mathrm{iso} I(\alpha, \delta)$, where $\phi^\mathrm{iso}$ corresponds to the isotropic flux (i.e., the flux averaged over the full celestial sphere), and $I(\alpha, \delta)$ is the {\it relative intensity} of the flux as a function of R.A. $\alpha$ and decl. $\delta$ in celestial coordinates.
Given that cosmic rays have been observed to be mainly isotropic, the flux is dominated by the isotropic term and therefore the {\it anisotropy} $\delta I = I-1$ is small.

The relative intensity gives the amplitude of deviations in the number of counts $N_\mathfrak{a}$ from the isotropic
expectation $\langle N_\mathfrak{a} \rangle$ in each angular bin $\mathfrak{a}$. The residual anisotropy $\delta I$ of the distribution of arrival directions of the cosmic rays is calculated by subtracting a reference map that describes the detector response to an isotropic flux
\begin{equation}
\delta I_\mathfrak{a} =  \frac{N_\mathfrak{a}- \langle N_\mathfrak{a}  \rangle}{\langle N_\mathfrak{a}  \rangle}~.
\label{eq:ri}
\end{equation}
In order to produce this reference map, we must have a description of the arrival direction distribution if the cosmic rays arrived isotropically at Earth. 

Ground-based experiments observe cosmic rays indirectly by detecting the secondary air shower particles produced by collisions of the cosmic-ray primary in the atmosphere. 
The observed large-scale anisotropy has an amplitude of about $10^{-3}$ but our simulations are not sufficiently accurate to describe the detector response at this level. We therefore calculate this expected flux from the data themselves in order to account for detector dependent rate variations in both time and viewing angle.
For Earth-based observatories, such a method requires averaging along each decl. band, thus washing out the vertical dependency (i.e. as a function of decl. $\delta$) in the relative intensity map $\delta I_\mathfrak{a}$.

A common approach is to estimate the relative intensity and detector exposure simultaneously using time-integration methods~\citep{Alexandreas1993,Atkins:2003ep}. 
However, these methods can lead to an under- or overestimation of the isotropic reference level 
for detectors located at mid latitudes,  since a fixed position on the celestial sphere is only observable over a relatively short period every day. 
As a result, the total number of cosmic ray events from this fixed position can only be compared against reference data observed during the same period. 
Therefore, time-integration methods can strongly attenuate large-scale structures exceeding the size of the instantaneous field of view~\citep{0004-637X-823-1-10}. 

\subsection{Maximum Likelihood Method}
For this analysis, we have relied on the likelihood-based reconstruction described in~\cite{0004-637X-823-1-10} and recently applied in the study of the large-scale cosmic-ray anisotropy by HAWC~\citep{hawclsa2017}. 
The method does not rely on detector simulations and provides an optimal anisotropy reconstruction and the recovery of the large-scale anisotropy projected on to the equatorial plane for ground-based cosmic ray observatories located in the middle latitudes as HAWC. 
The generalization of the maximum likelihood method for combined data sets from multiple observatories that have exposure to overlapping regions of the sky is described in Appendix~\ref{sec:lh}.

\subsection{Statistical Significance}
In order to calculate the statistical significance of anisotropy features in the final reconstructed map, \cite{0004-637X-823-1-10} generalizes the method in \cite{Li:1983fv} to account for the optimization process of the time-dependent exposure.
The significance map (in units of Gaussian $\sigma$) is then calculated as
\begin{equation}\label{eq:significance}
  S_{ i } = \sqrt{2}\left(-\mu_{ i ,\rm on}+\mu_{ i ,\rm off} + n_ i \log\frac{\mu_{ i ,\rm on}}{\mu_{ i ,\rm off}}\right)^{1/2}\,.
\end{equation}
For each pixel $i$ 
in the celestial sky, we define expected {\it on-source} and {\it off-source} event counts from neighbor pixels in a disc of radius $r$ centered on that pixel. For this analysis we have chosen a radius of 5$^\circ$.  Given the set of pixels $\mathcal{D}_ i $, the observed and expected counts are
\begin{align}
  n_{ i } &= \sum_{ j \in\mathcal{D}_{ i }}\sum_{\tau}n_{\tau  j }\,,\\
  \mu_{ i ,\rm on} &=  \sum_{ j \in\mathcal{D}_{ i }}\sum_{\tau}\mathcal{A}_{\tau j }\mathcal{N}_\tau I_{ j }\,,\\
  \mu_{ i ,\rm off} &= \sum_{ j \in\mathcal{D}_{ i }}\sum_{\tau}\mathcal{A}_{\tau j }\mathcal{N}_\tau I^{\rm reference}_{ j }\,,
\end{align}
where $\mathcal{A}_{\tau j}$ is the relative acceptance of the detector in pixel $j$ and sidereal time bin $\tau$, $\mathcal{N}_\tau$ gives the expected number of isotropic events in sidereal time bin $\tau$, $I_{j}$ is the relative intensity, and where $I=I^{\rm reference}+I^{\rm residual}$ is divided into a contribution from the reference map and the residual relative intensity. 
For small-scale features, $I^{\rm reference}$ corresponds to the first 3 spherical harmonic components ($\ell \leq 3$) of the relative intensity. 
In order to distinguish excess and deficit, we multiply Eq.~\ref{eq:significance} by the sign of each smoothed pixel $\delta I_i$ in the anisotropy map.

\subsection{Harmonic Analysis and Dipole Fit}
The relative intensity can be decomposed as a sum over spherical harmonics $Y^{\ell m}$,
\begin{equation}
\label{eq:sphHarmonics}
  {\delta I(\mathbf{\mathbf{u}}_i)} =
    \sum_{\ell=1}^{\infty}
    \sum_{m=-\ell}^{\ell} a_{\ell m} Y_{\ell m}(\mathbf{u}_i)\,.
\end{equation}
The vector components of the dipole in terms of the spherical harmonic expansion $Y_{\ell m}$ in equatorial coordinates are related to the
$a_{\ell m}$ coefficients with
\begin{equation}\label{eq:ahlers1}
\boldsymbol{\delta} 
\equiv  (\delta_{0h}, \delta_{6h},\delta_{N}) = \sqrt{\frac{3}{2\pi}}(-\Re(a_{11}), \Im(a_{11}),a_{10}) \,,
\end{equation}
where $\Re(a_{11})$ and $\Im(a_{11})$ are respectively, the real and imaginary components of $a_{11}$, and taking into account that $a_{1-1} = -a^*_{11}$ and $a_{10} = a^*_{10}$ (see~\cite{Ahlers:2016rox}). 

From equation \ref{eq:ahlers1} and the $a_{\ell m}$ coefficients, 
one can obtain the horizontal components of the dipole 
$\delta_{0h}$ and $\delta_{6h}$ with respect to the $0$h and $6$h R.A. axes.
The phase and
amplitude of the projected dipole on the equatorial plane are given by 
\begin{equation}\label{eq:ahlers2}
 (\delta_{0h}, \delta_{6h}) = (\tilde{A}_1 \cos \alpha_1, \tilde{A}_1 \sin \alpha_1) \,,
\end{equation}
where
$\alpha_1$ is the phase and $\tilde{A}_1$ is the amplitude of the projected  dipole on the equatorial plane and it is related to the true amplitude $A_1$ through the dipole inclination $\delta_0$ with $\tilde{A}_1 = A_1 \cos \delta_0$.
\begin{figure*}[ht]
\begin{center}
\includegraphics[width=0.65\textwidth]{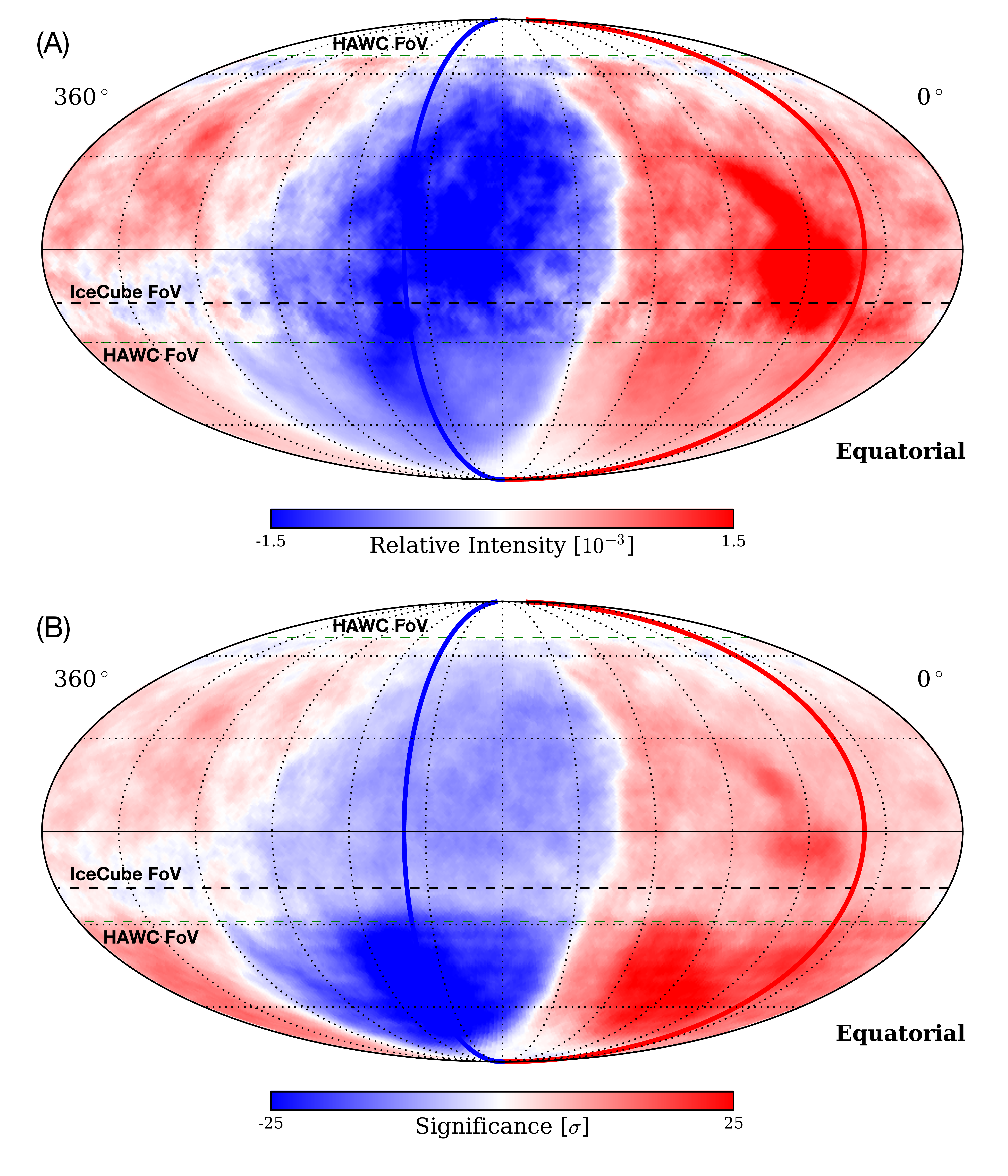}
              \caption{Mollweide projection sky maps of (A) relative intensity $\delta I_\mathfrak{a}$ (Eq.~\ref{eq:ri}) of cosmic-rays at 10 TeV median energy 
and (B) corresponding 
signed statistical significance $S_{i}$ (Eq. \ref{eq:significance}) of the deviation from the average intensity in J2000 equatorial coordinates.
The thick red and blue lines in the figures indicate correspondingly, the node and antinode of the phase in R.A. of the dipole component from the fit.}
        \label{fig:map:ls-comb}
        \end{center}
\end{figure*}

\subsection{Angular Power Spectrum}\label{sec:power_spectrum}
The angular power spectrum for the relative intensity field is defined as: 
\begin{equation}
  \mathcal{C}_{\ell} = \frac{1}{2 \ell + 1} \sum_{m=-\ell}^{\ell} | a_{\ell m} |^{2}~~,
   \label{eq:cldef_true}
\end{equation}
for each value of $\ell$.
Since this analysis is not sensitive to the vertical component of the anisotropy, the largest recoverable dipole amplitude $\tilde{A}$ has the terms $m = 0$ missing and we can only measure a pseudo power spectrum $\tilde{\mathcal{C}}_{\ell}$:
\begin{equation}
  \mathcal{\tilde{C}}_{\ell} = \frac{1}{2 \ell} \sum_{m=-\ell, m\neq 0}^{\ell} | a_{\ell m} |^{2}~~,
   \label{eq:cldef}
\end{equation}
The angular power spectrum provides an estimate of the significance of structures at different angular scales of $\sim$ 180$^\circ/\ell$. 
In the ideal case of a $4\pi$ sky coverage, the multipole moments ${a}_{\ell m}$ of the reconstructed anisotropy would carry all the information of the anisotropy (except for the $m=0$ vertical component terms). However, as will be discussed in Section \ref{sec:systematics}, partial sky coverage of individual experiments further limits the amount of information that can be obtained  from the reconstructed pseudo multipole moment spectrum. 

\section{Results}\label{sec:results}

The measured relative intensity map is shown in Figure~\ref{fig:map:ls-comb}. 
A  smoothing procedure was applied to all maps using a top-hat function in which a single pixel's value is the average of all pixels within a 5$^\circ$ radius.
The map shows an anisotropy in the distribution of arrival directions of 
cosmic rays with 10 TeV median primary particle energy that extends across both hemispheres. 
The significance (smoothed by summing over pixels) of the IceCube region reflects the much larger statistics in the IceCube dataset compared to that from HAWC at energies of $\sim$10 TeV. 
\begin{figure*}[ht]
\begin{center}
\includegraphics[width=0.65\textwidth]{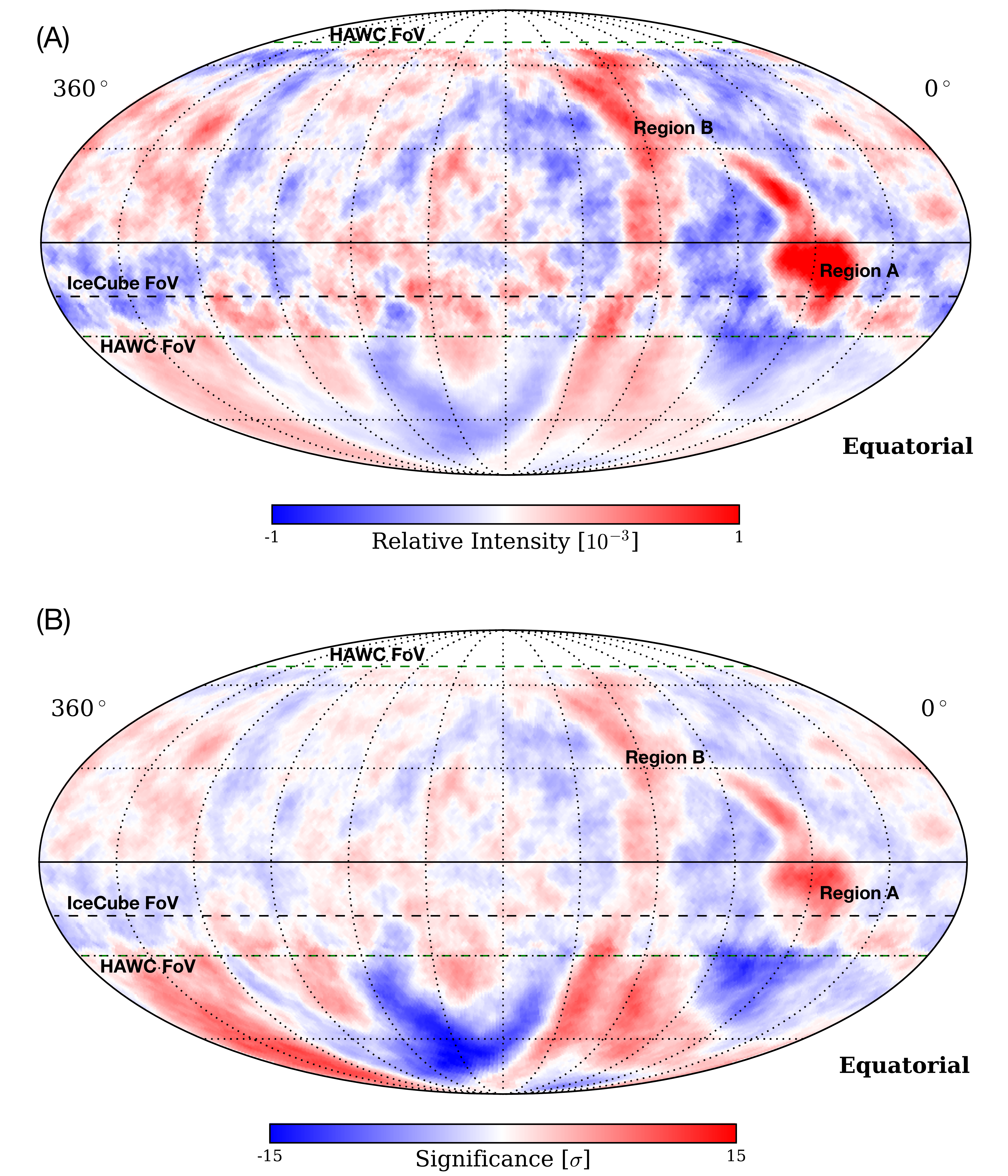}
\caption{(A) Relative intensity $\delta I_\mathfrak{a}$ (Eq.~\ref{eq:ri})
after subtracting the multipole fit from the large-scale map and (B) corresponding 
signed statistical significance $S_{i}$ (Eq. \ref{eq:significance})  of the deviation from the average intensity in J2000 equatorial coordinates.}
        \label{fig:small:relint}
\end{center}
\end{figure*}

\begin{figure*}[ht]
\centering
\includegraphics[width=.7\textwidth]{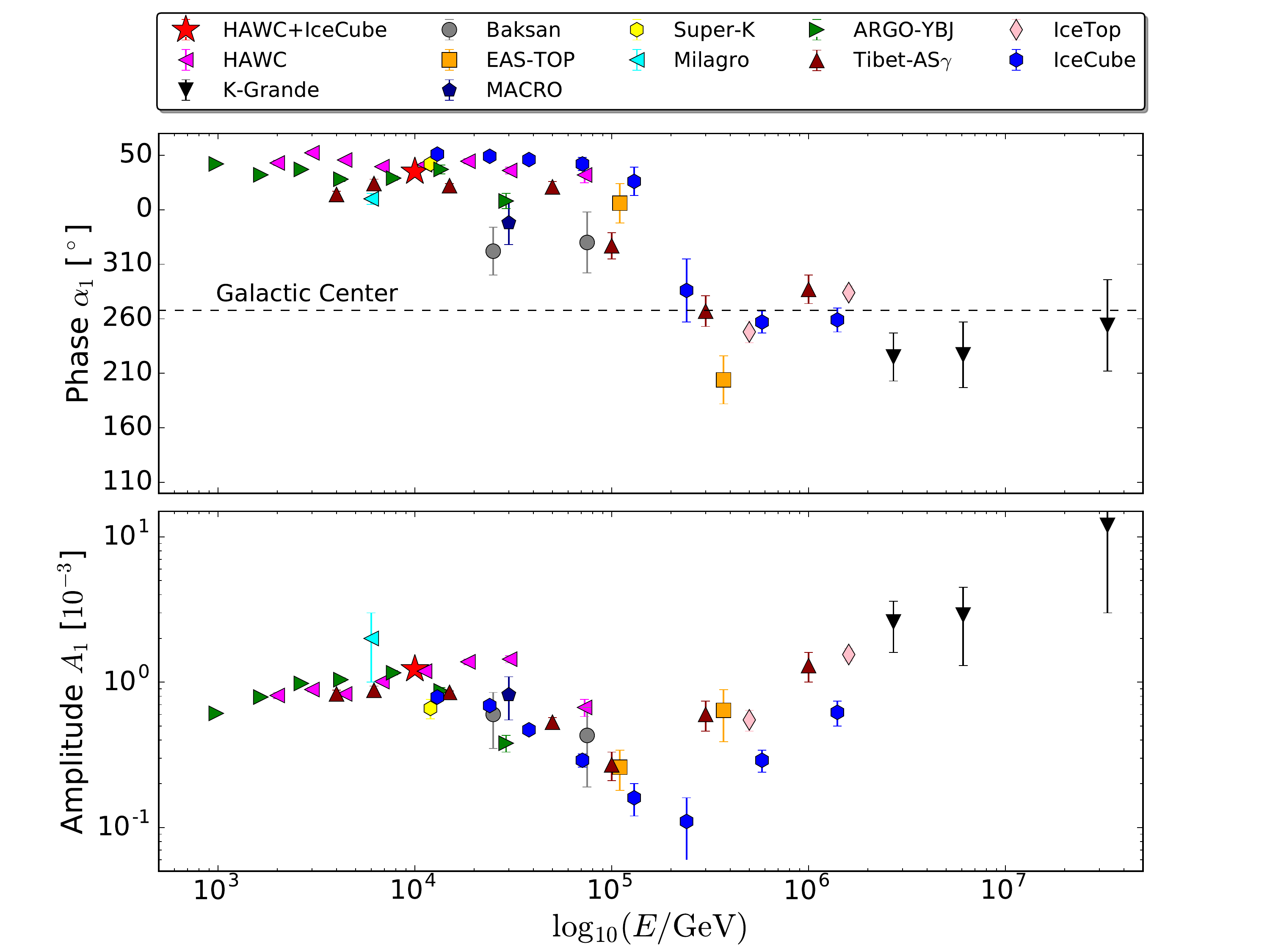}
\caption{Reconstructed dipole component amplitude and phase from this measurement along previously published TeV-PeV results from other experiments (adopted from~\cite{Ahlers:2016rox}). The  results shown are from~\cite{hawclsa2017,kgrande2015,baksan2009,aglietta_2009,macro2003,guillian_2007,abdo_2009,bartoli_2015,amenomori_2005,IceCube:2013mar,icecube2016}}
\label{fig:phase}
\end{figure*}

Figure~\ref{fig:small:relint} is the residual small-scale anisotropy after subtracting the fitted multipole from the spherical harmonic expansion
with $\ell \leq 3$ 
from the large-scale map in Figure~\ref{fig:map:ls-comb} in order to reveal structures smaller than $60^\circ$.
The large-scale structure and significant small-scale structures in Figures~\ref{fig:map:ls-comb} and~\ref{fig:small:relint} are largely consistent with previous individual measurements, as shown in Figure~\ref{fig:phase}.
Observed features extend across the horizon of both datasets. The one referred to as ``region A'' by the Milagro Collaboration~\citep{Milagro:2008nov} roughly extends from $(54^\circ,  -16^\circ)$, to $(78^\circ, 18^\circ)$, in equatorial coordinates ($\delta$, $\alpha$). The so called ``region B''~\citep{Milagro:2008nov} corresponds to the boundary between the excess and deficit regions (see Figure \ref{fig:map:ls-comb}) in the northern sky that appears as a small scale feature (see Figure \ref{fig:small:relint}) for short integration times.

\begin{table*}[ht]
\caption{Spherical Harmonic Coefficients [$10^{-4}$]}
\begin{center}
\footnotesize{
\begin{tabular}{l|lllllllll}
\hline
$\ell$ &$m=$1&2&3&4&5&6&7\\
\hline \\
1 & -13.26 +10.49$i$\\
2 & -0.21 -3.6$i$& -7.20 +2.05$i$\\
3 & { }1.75 -1.7$i$& -2.03 +0.13$i$& { }0.20 +0.17$i$\\
4 & { }1.70 -0.52$i$& { }0.07 +1.69$i$& -0.86 -0.8$i$& -1.19 +0.04$i$\\
5 & { }0.58 +0.27$i$& -0.07 -1.1$i$& -1.64 -0.051$i$& { }0.18 -0.15$i$& -0.11 -1.5$i$\\
6 & { }0.80 -0.88$i$& -0.24 -0.38$i$& -0.10 +0.63$i$& { }0.13 -1.2$i$& { }0.27 +0.47$i$& { }1.65 -0.53$i$\\
7 & { }0.44 -0.67$i$& { }0.37 +0.15$i$& -0.21 -0.14$i$& -0.70 +0.04$i$& { }0.84 -0.27$i$& { }0.13 -0.54$i$& { }0.07 +0.91$i$\\
8 & { }0.26 +0.06$i$& { }0.14 -0.47$i$& -0.39 -0.22$i$& -0.42 +0.72$i$& -0.15 -0.15$i$& -0.72 -0.61$i$& { }0.42 +0.36$i$\\
9 & { }0.11 -0.88$i$& -0.29 -1.3$i$& { }0.22 -0.17$i$& { }0.12 -0.56$i$& -0.01 -0.34$i$& { }0.60 +0.47$i$& -0.06 -0.48$i$\\
10 & { }0.21 -0.97$i$& { }0.25 -0.5$i$& { }0.21 -0.65$i$& { }0.09 -0.088$i$& -0.10 +0.12$i$& { }0.11 -0.017$i$& { }0.02 +0.19$i$\\
11 & { }0.56 -0.39$i$& { }0.06 -0.42$i$& -0.15 -0.68$i$& -0.04 +0.05$i$& -0.26 +0.04$i$& -0.07 -0.26$i$& -0.16 +0.25$i$\\
12 & { }0.40 +0.07$i$& { }0.19 -0.56$i$& -0.27 -0.48$i$& -0.17 -0.1$i$& -0.13 -0.18$i$& -0.03 -0.23$i$& { }0.33 +0.13$i$\\
13 & { }0.45 -0.33$i$& -0.04 -0.69$i$& { }0.17 -0.92$i$& -0.26 -0.6$i$& { }0.13 +0.24$i$& -0.08 +0.02$i$& { }0.04 +0.04$i$\\
14 & { }0.57 -0.16$i$& { }0.13 -0.53$i$& { }0.17 -1.1$i$& -0.31 -0.089$i$& { }0.08 -0.09$i$& -0.25 -0.12$i$& -0.05 +0.22$i$\\
\hline \\
$\ell$ &$m=$8&9&10&11&12&13&14\\
\hline \\
8 & -0.54 +0.19$i$\\
9 & { }0.15 +0.64$i$& -0.04 +0.45$i$\\
10 & { }0.22 +0.12$i$& -0.66 -0.57$i$& -0.26 +0.38$i$\\
11 & { }0.25 +0.02$i$& -0.21 -0.4$i$& { }0.15 -0.25$i$& -0.06 -0.18$i$\\
12 & { }0.37 +0.09$i$& -0.46 +0.25$i$& -0.13 +0.20$i$& -0.08 +0.21$i$& { }0.04 -0.18$i$\\
13 & { }0.11 +0.13$i$& { }0.13 -0.13$i$& -0.35 -0.098$i$& { }0.39 +0.45$i$& -0.01 -0.3$i$& { }0.41 -0.17$i$\\
14 & -0.13 +0.34$i$& { }0.36 -0.11$i$& -0.04 -0.072$i$& -0.11 -0.17$i$& -0.19 +0.32$i$& { }0.13 +0.21$i$& { }0.18 +0.35$i$\\
\hline \\
\end{tabular}
}
\end{center}
\label{tab:alms}
\end{table*}
We obtain the $ a_{\ell m}$ through a transformation of spherical harmonics using the {\tt HEALPix} function \emph{map2alm}. The results are presented in  Table~\ref{tab:alms}.
The horizontal components of the dipole obtained from equation
(\ref{eq:ahlers1}) using the $a_{\ell m}$ values in Table \ref{tab:alms} are
$\delta_{0h} =  9.16 \times 10^{-4}$ and
$\delta_{6h} = 7.25 \times 10^{-4}~(\pm0.04 \times 10^{-4})$, respectively, with respect to the $0$h and $6$h R.A. axes.
The dipole amplitude and phase
$\tilde{A}_1=(1.17\pm.01) \times 10^{-3}$, 
$\alpha_1=38.4\pm0^\circ.3$ measured in this combined study are shown in Figure~\ref{fig:phase} along with
previously published results from other experiments in the TeV-PeV primary particle energy range. The combined systematic uncertainty in the amplitude and phase of the dipole are expected to be $\delta \tilde{A}_1\sim 0.06 \times 10^{-3}$, and $\delta \alpha_1 \sim 2^\circ.6$ respectively (see section~\ref{sec:syst}).

\begin{figure*}[ht]
\begin{center}
\includegraphics[width=0.7\textwidth]{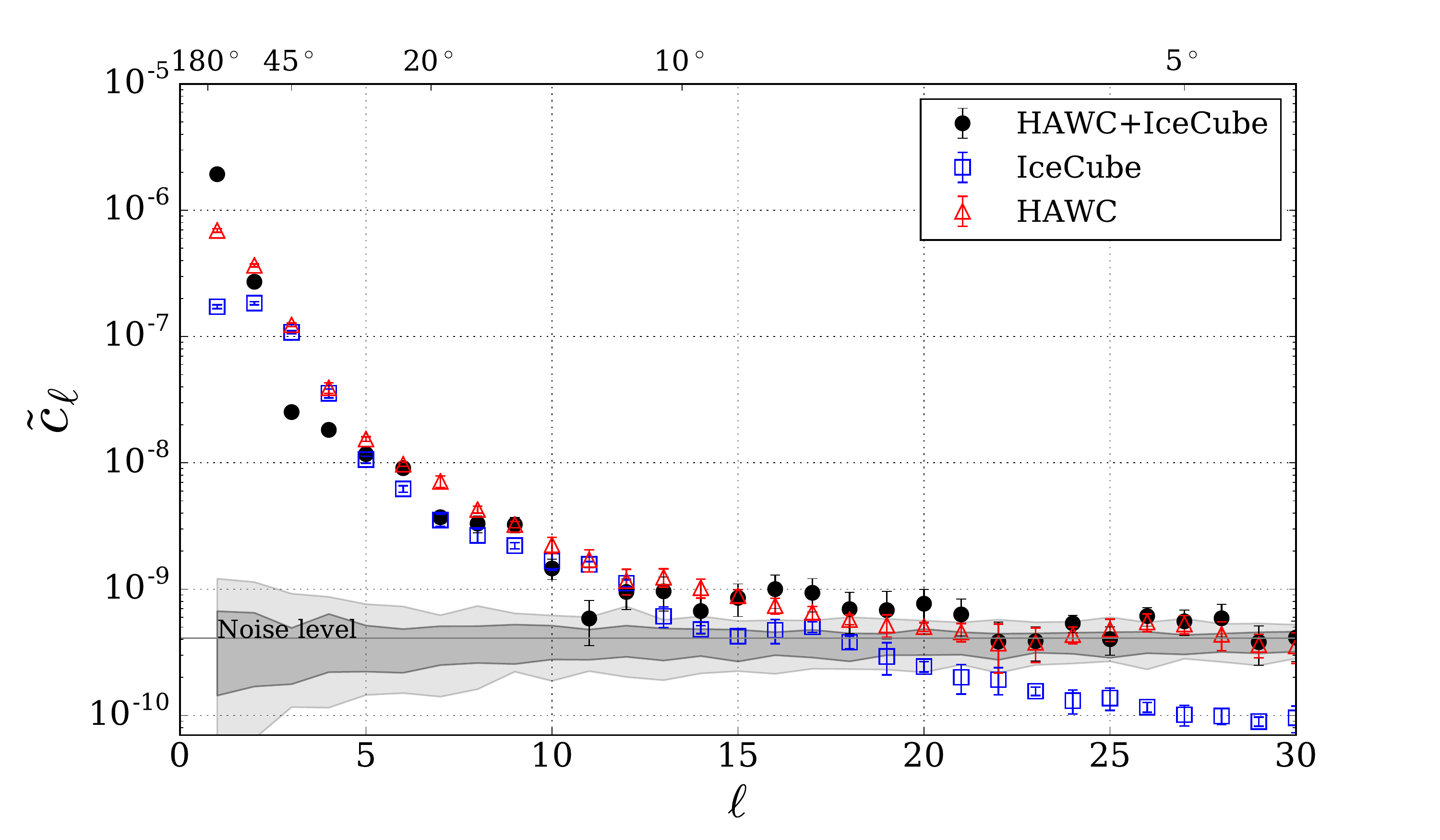}
\caption{Angular power spectrum of the cosmic ray anisotropy at 10 TeV.
The gray band represents the 90\% confidence level around the level of statistical fluctuations for isotropic sky maps.
The noise level is dominated by limited statistics for the portion of the sky observed by HAWC. The IceCube dataset alone has a lower noise level and is sensitive to higher $\ell$ components. 
The dark and light gray bands represent the power spectra for isotropic sky maps at the 68\% and 95\% confidence levels respectively.
The errors do not include systematic uncertainties from partial sky coverage.
}
\label{fig:power-spectrum}
\end{center}
\end{figure*}
The angular power spectrum for the combined dataset in Figure~\ref{fig:power-spectrum} provides an estimate of the significance of structures at different angular scales of $\sim$ 180$^\circ/\ell$. Biases are substantially reduced with the likelihood method and by eliminating degeneracy between multipole moments with a nearly full sky coverage. The angular power spectrum can therefore be considered to be the physics fingerprint of the observed 10 TeV anisotropy, providing information about the propagation of cosmic rays and the turbulent nature of the Local Interstellar Magnetic Field (LIMF)~\citep{Giacinti:2011mz,Ahlers:2016rox}. The large discrepancy between the combined and individual datasets is the result of the limited sky coverage by each experiment. This systematic effect will be discussed in Section \ref{sec:systematics}.
A residual limitation in this analysis is the fact that ground-based experiments are generally not sensitive to the vertical component of the anisotropy as discussed by~\cite{hawclsa2017} and~\cite{0004-637X-823-1-10}, as mentioned earlier.

The measured quadrupole component has an amplitude of $6.8 \times 10^{-4}$ and is inclined at $20.7~\pm0^\circ.3$ above (and below) equatorial plane. 
As with the dipole, the fitted quadrupole component from the spherical harmonic expansion is also missing the $m = 0$ terms. However, the combination of $a_{21}$ and $a_{22}$ non-vertical quadrupole components can still provide valuable information. 
The experimental determination of the vertical components of the anisotropy would require accuracies better than the amplitude of the anisotropy ($\sim 10^{-3}$). 
This becomes easier at ultra-high energies where a dipole of much larger amplitude has been observed~\citep{auger1266}. 
The full-sky coverage also provides better constraints for fitting the $\ell = 2$ and $\ell = 3$ multipole components and reduces correlations  between spherical harmonic expansion coefficients $a_{\ell m}$.

\section{Systematics Studies}
\label{sec:syst}
\subsection{Overlapping Region}
\begin{figure}[h]
\begin{center}
\includegraphics*[width=0.99\columnwidth]
{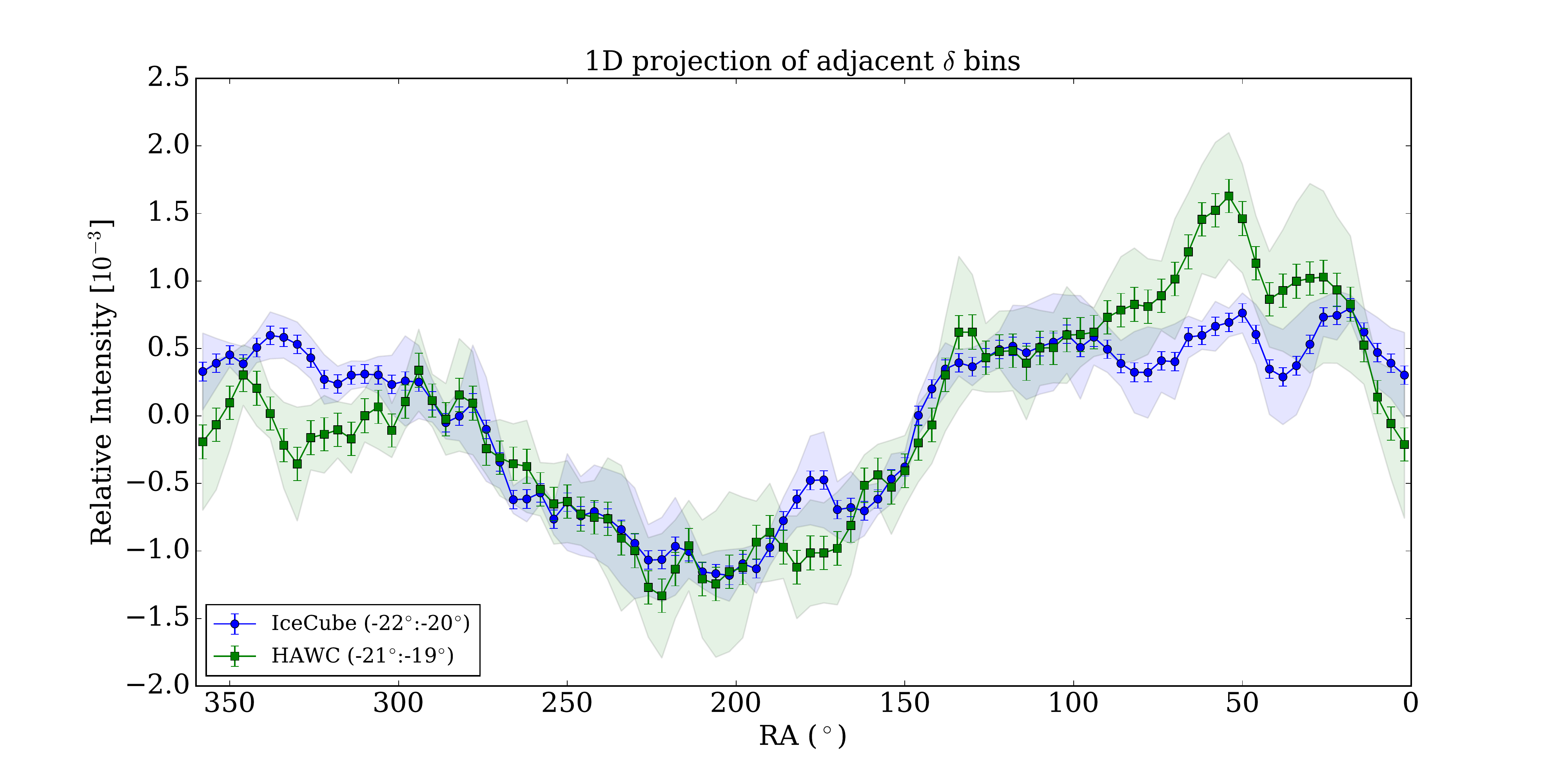} \end{center}
\vskip -0.7cm
\caption{\small
One-dimensional R.A. projection of the relative intensity of cosmic rays 
for adjacent $\delta$ bins in the overlap region at -20$^\circ$
for HAWC and IceCube data. 
There is general agreement for large scale structures. 
The two curves correspond to different $\delta$ bands. The shaded bands correspond to systematic uncertainties due to mis-reconstructed events,
derived from the relative intensity distributions in adjacent decl. bands between $-25^{\circ}$ and $-15^{\circ}$.
} \label{fig:overlap}
\end{figure}
We have studied two adjacent $\delta$ bands at -20$^\circ$ for HAWC and IceCube data near the horizon of each detector (see Figure~\ref{fig:overlap}). The HAWC band extends from -21$^\circ$ to -19$^\circ$ while the IceCube band extends from -22$^\circ$ to -20$^\circ$. 
The large structure between the two datasets is consistent though small structures differ. 
It is worth noting that the overlap region is where we expect to find the largest difference in median energy between the two datasets (see Figure~\ref{fig:energy_dec}).
The angular resolution of both detectors also decreases toward the horizon. While HAWC data has a smaller point spread function at this decl. and is sensitive to structures on smaller scales, IceCube has better statistics so the structures are more significant. 
One particular feature that stands out is the excess in HAWC around $\alpha=50^\circ$ that coincides with the so called ``region A''. 
There appears to be a corresponding small excess in the IceCube data.    
It is also worth noting that statistics in this region are quickly decreasing with increasing zenith angle as is the quality of angular reconstructions. As a result, $\delta$ bins closer to the horizon contain a high level of contamination from bins in higher zenith angles.

\subsection{Partial Sky Coverage}\label{sec:systematics}
\begin{figure}[ht]
\begin{center}
\includegraphics[width=0.99\columnwidth]{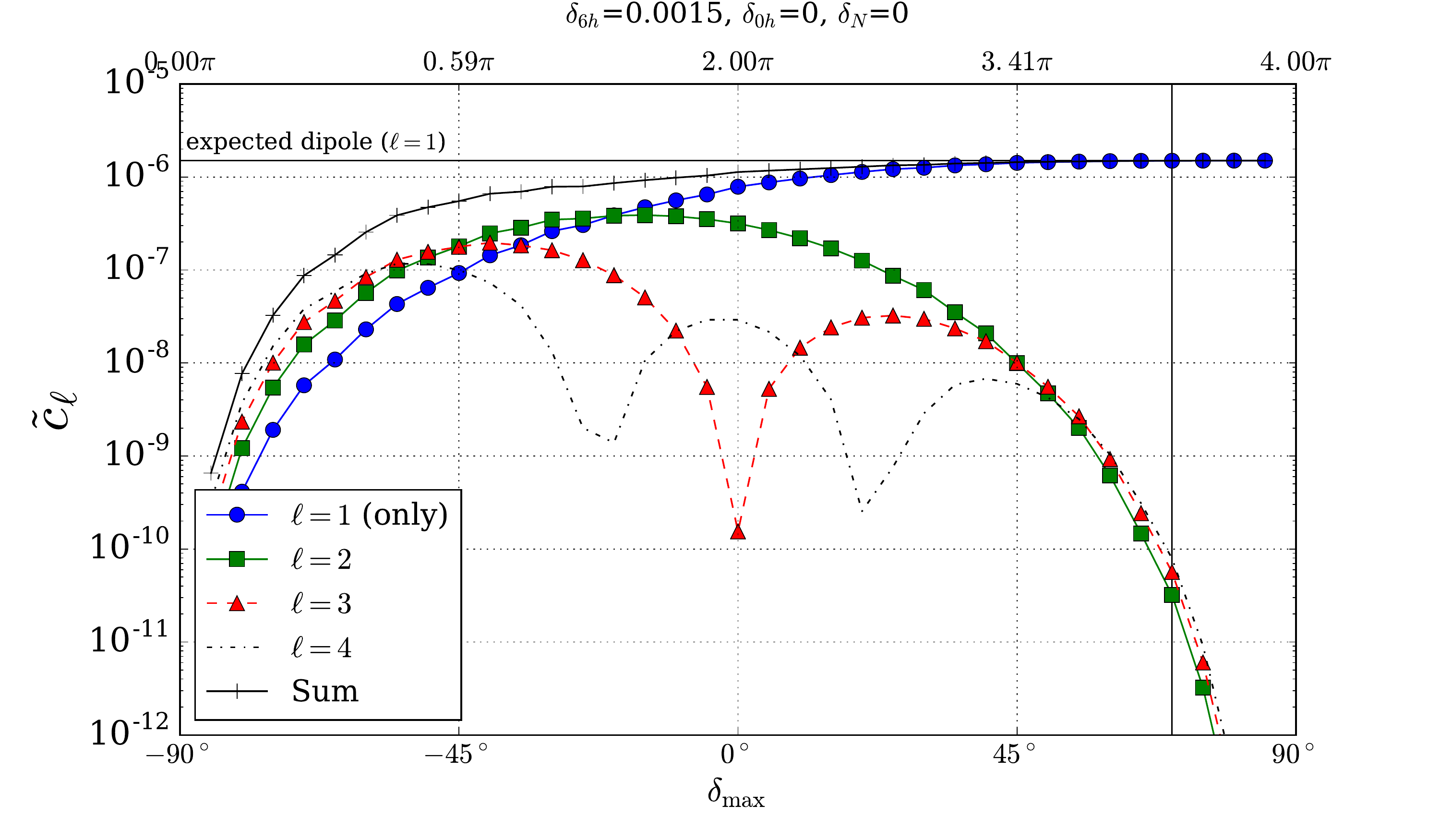}

              \caption{
 Angular power spectrum as a function of sky coverage for $\ell = \{1, 2, 3, 4\}$. The horizontal axis indicates the maximum decl. $\delta_\mathrm{max}$, keeping $\delta_\mathrm{min} = -90^\circ$ for a dipole injected horizontally in direction $\delta_{6h}$. The partial coverage of sky produces an artificial quadrupole and octupole that decrease in power with greater celestial coverage.           
              }
        \label{fig:toy:horiz}
\end{center}
\end{figure}
Incomplete coverage of the sky leads to an underestimation of the angular power of the dipole perpendicular to the axis of rotation of the Earth. 
The pseudo-moments of the projected dipole, $a_{11}$ and $a_{1-1}$, are corrected by a geometric factor introduced by~\cite{0004-637X-823-1-10} 
in order to estimate the true moments $\hat{a}_{11}$ and $ \hat{a}_{1-1}$. 
Furthermore, there is a degeneracy between different $\ell$ pseudo-modes under partial sky coverage that primarily affects the multipolar components $\ell = 2$, $\ell = 3$, and to a lesser degree, $\ell = 4$ as has been previously studied by~\cite{SOMMERS2001271}.
This effect is evident in Figure~\ref{fig:toy:horiz} which corresponds to a dipole injected horizontally in the direction $\delta_\mathrm{6h}$. The partial coverage of the sky produces an artificial quadrupole, octupole and hexadecapole that, in the case of a horizontal dipole, decrease in power with greater celestial coverage. The horizontal axis indicates the maximum observable decl. $\delta_\mathrm{max}$, keeping $\delta_\mathrm{min} = - 90^\circ$.
\begin{figure*}[ht]
\begin{center}
\includegraphics[width=0.65\textwidth]{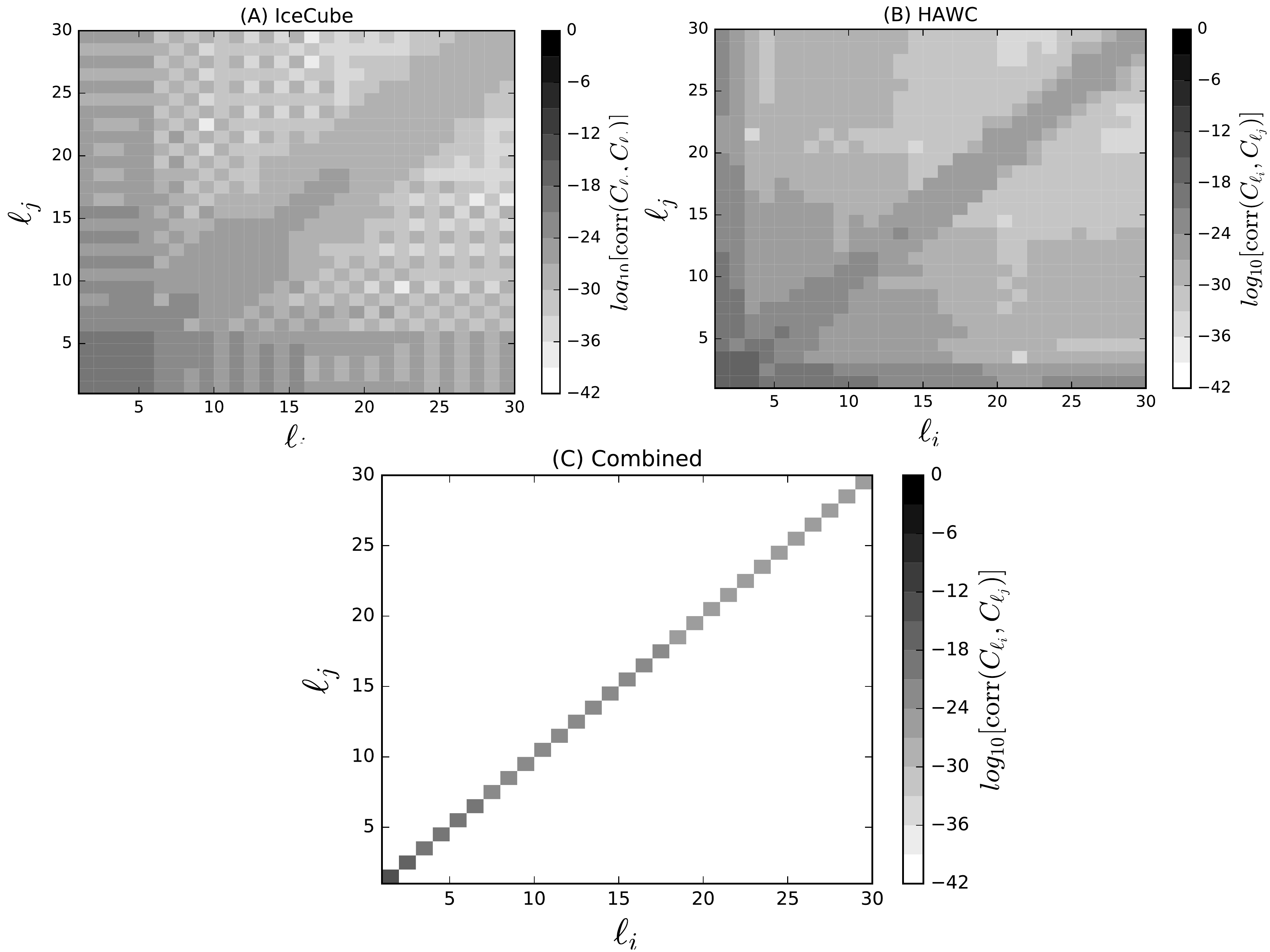}
        \caption{Correlation matrix for $C_\ell$ modes with partial sky coverage from individual experiments (A, B) and for the combined field of view (C).}
        \label{fig:correlation_matrix}
\end{center}
\end{figure*}
From Figure~\ref{fig:toy:horiz} it is possible to see that the spurious quadrupole and octupole components (which are significant for partial integrated sky coverage) are reduced to an amplitude to order $10^{-5}$ in this analysis.
Figure~\ref{fig:correlation_matrix} shows the correlation matrix~\citep{efstahiou2004} of the different $\ell$-modes up to $\ell=30$ calculated using the {\tt PolSpice}\footnote{
PolSpice website: http://www2.iap.fr/users/hivon/software/PolSpice/.} software package. The correlation between $\ell$-modes due to partial sky coverage is appreciable for larger $\ell$, though to a lesser degree.

\subsection{Seasonal Variations and Local Variations in Solar Time}
The relative motion of the Earth around the Sun can introduce a systematic solar dipole, a dipole anisotropy analogous to the Compton-Getting effect~\citep{Compton:1935} produced by the motion of Earth around the Sun, that points in the direction of Earth's orbital velocity vector. 
The influence of diurnal variations (such as the solar dipole) on the sidereal anisotropy can be estimated from the influence it has on the anti-sidereal distribution in a frame with 364.24 cycles per year (see, e.g.~\cite{guillian_2007}). Any significant variations in this frame result from a modulation of the solar frame and represents a systematic effect of the solar frame on the sidereal anisotropy~\citep{icecube2016}. The anti-sidereal distribution of the HAWC dataset has a maximum amplitude of $5 \times 10^{-5}$. Both contamination from the solar dipole and atmospheric pressure variations are included in this systematic.
For IceCube, the same systematic uncertainty is at the level of $\sim 3\times 10^{-5}$. The worst-case uncertainty on the reconstructed phase of the dipole is $\delta \alpha = 2^\circ.6$ and a combined systematic uncertainty of $\delta \tilde{A} = 6 \times 10^{-5}$ for the dipole amplitude. 

The solar dipole anisotropy produced by the motion of Earth around the Sun is given by the equation
\begin{equation}\label{eq:solardipole}
\frac {\Delta I} {I} = (\gamma + 2) \frac {v} {c} \cos(\theta_v) \,,
\end{equation}
where $I$ is the cosmic-ray intensity, $\gamma$ is the index of the differential energy spectrum of cosmic rays, $v$ is the velocity of the Earth, $c$ is the speed of light 
and  $\theta_v $ is the angle between the direction of the reconstructed cosmic rays and the direction of the velocity vector~\citep{Compton:1935}.
This vector rotates by $360^\circ$ such that, after one year, the effect is ideally completely cancelled for 100\% duty cycle of observation. However, a residual dipole can be introduced if the data does not cover an integer number of years with uniform coverage. 
In other words, any gaps in data taking can result in a slight bias to the measured dipole. A solar dipole anisotropy at the level of $10^{-4}$ has been previously observed at 
several TeVs~\citep{Amenomori:2004bf,Amenomori:2006bx,abdo_2009,Abbasi:2011ai,IceCube:2012feb,bartoli_2015}. Based on Monte Carlo studies, the residual contribution solar dipole that results from gaps in data taking is estimated to be of order $\sim 10^{-5}$ for the HAWC dataset, which is smaller than the statistical error of this analysis. In the case of IceCube, the detector has an uptime of 99\% (see ~\cite{IceCube:detector})  reduced to an uptime of 95.4\% after selecting full sidereal days. As a result, the systematic effect of data gaps is smaller~\citep{IceCube:2012feb}. 

In addition to variations caused by the anisotropy and the solar dipole, there may also be local variations in the detection of cosmic rays caused by changes in atmospheric conditions, such as pressure and temperature, and also by changes in the detector. 
For 10 TeV energies, HAWC is located below the shower maximum X$_\mathrm{max}$ for all primary masses. As a result, an increase in pressure leads to an increase of the atmospheric overburden which results in an attenuation of shower sizes. Atmospheric overburden is related to ground pressure $p$ as $X_0 = p/g$, where $g =$ 9.87~m~s$^{-2}$ is the local gravitational acceleration~\citep{ABBASI201340}.
In first order approximation, the simple correlation
between the change in the logarithm of the rate $\Delta\{\ln R \}$ and the surface
pressure change $\Delta P$ is
\begin{equation}\label{eq:pressurecorr}
\Delta\{\ln R\} = \beta \cdot \Delta P
\end{equation}
where $\beta$ is the barometric coefficient~\citep{Tilav:2010jan}.
The variations in atmospheric pressure at the HAWC site are primarily due to \emph{atmospheric tides} driven by temperature and a small contribution from \emph{gravitational tides}~\citep{JGRA:JGRA20326}. 
We have studied the effect of atmospheric pressure variations by applying a correction to the data rate to account for measured changes in pressure at the HAWC site. The procedure involves determining the correlation coefficient between the surface pressure data and the detector rate from Eq.~\ref{eq:pressurecorr} in order to weight individual events. This yields a barometric coefficient of $\beta = -0.0071$ hPa$^{-1}$ The residual contamination from atmospheric variations is estimated to be on the order of $\sim 10^{-6}$. Temperature variations in the stratosphere can introduce a similar effect with a 24h cycle and a 365 day cycle. However, this effect is small for latitudes near the equator and in the case of the daily variations, it is a smaller effect than that of pressure variations. 

In contrast with HAWC, where the event rate is anti-correlated with atmospheric pressure and with the effective temperature of the stratosphere, the muon rate in IceCube is directly correlated with the effective temperature~\citep{Tilav:2010jan}.
Event rate variations in IceCube have an annual period since one day at the South Pole lasts 365 days instead of 24 hours. In the case of IceCube there are also faster atmospheric variations of lower amplitude but these approximately affect the event rate globally in all azimuth directions (with a maximum Kolmogorov-Smirnov distance below $9.6 \times 10^{-4}$ for daily variations at a 90\% confidence level). Due to the geometry of the detector and its location at the South Pole, this also means that such variations equally affect every angle in R.A. 

Seasonal variations in the effective temperature can introduce modulations in the intensity of the Solar dipole. As a result, the Solar dipole would not average to zero over a full year and thus would produce a residual bias. However, the amplitude of the anti-sidereal distribution indicates that this is not a significant effect.

      

\section{Discussion}

The combined sky map of arrival direction distribution of the 10 TeV cosmic rays collected by HAWC and IceCube and the corresponding power spectrum of its spherical harmonics components, may provide important hints on the origin of the observation. In particular, the angular power spectrum can reveal information about how cosmic rays propagate through the interstellar medium while the large-scale arrival direction distribution provides hints about the structure of the nearby LIMF and the heliosphere.

\begin{figure*}[ht]
\begin{center}
\includegraphics[width=0.65\textwidth]{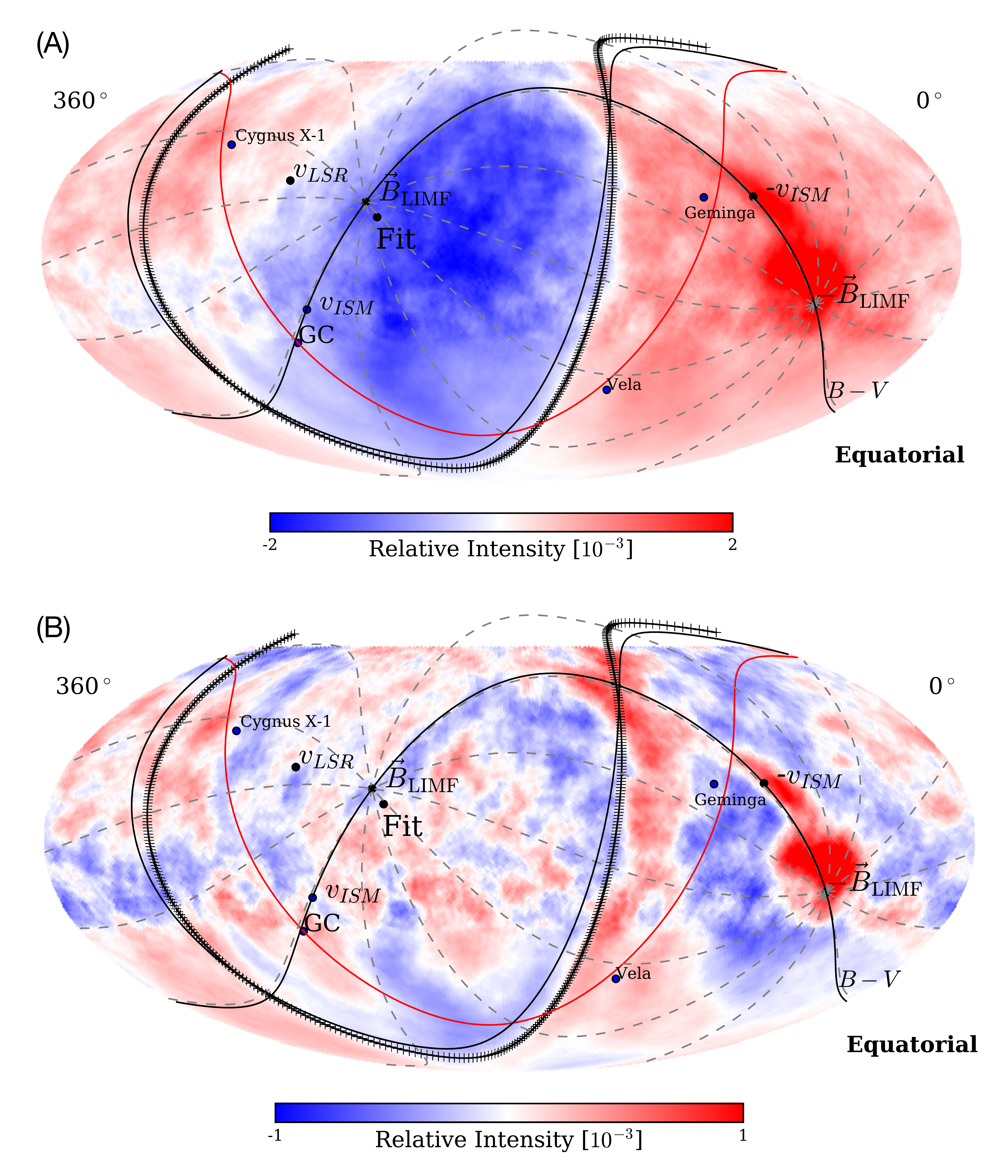}
 \caption{(A) Relative intensity of cosmic-rays at 10 TeV median energy (Figure~\ref{fig:map:ls-comb}(A)) and (B) corresponding small-scale anisotropy (Figure~\ref{fig:small:relint}(A)) in J2000 equatorial coordinates with color scale adjusted to emphasize features. The fit to the boundary between large scale excess and deficit regions is shown as a black crossed curve. The magnetic equator from~\cite{2041-8205-818-1-L18} is shown as a black curve as is the plane containing the local interstellar medium magnetic field and velocity ($B-V$ plane). The Galactic plane is shown as a red curve and two nearby supernova remnants, Geminga and Vela are shown for reference as is Cygnus X-1, a black hole X-ray binary known to produce high energy 
$\gamma$ rays~\citep{1538-4357-665-1-L51}.}
        \label{fig:map:comb_feat}
        \end{center}
\end{figure*}
\subsection{Cosmic Ray Propagation in the Interstellar Medium}
The angular power spectrum in Figure~\ref{fig:power-spectrum} shows two different regimes: a steeply falling slope at large scales $\ell =1, 2, 3$ and a softer slope at small scales $\ell > 3$. This suggests that two different mechanisms are responsible for the structures observed in the sky map.
The steep portion of the angular power spectrum may be associated with large scale diffusive processes (over many mean free paths) across the interstellar medium, as suggested by~\cite{erlykin_2006,Blasi:2012jan,Ptuskin:2012dec,pohl_2013,sveshnikova_2013,savchenko_2015,ahlers_2016,0004-637X-835-2-258}.
On the other hand, the softer slope portion appears to be consistent with non-diffusive pitch angle scattering effects on magnetic turbulence within the mean free path~\citep{Giacinti:2011mz} and with that obtained from numerical calculations of sub-PeV protons propagating through incompressible magnetohydrodynamic turbulence~\citep{barquero2016}.
In~\cite{Ahlers2014} it is shown that under certain conditions, those small-scale structures arise as natural consequence of hierarchical evolution of angular scales under Liouville's theorem.

The dipole component of the anisotropy may provide a hint into the direction of the large scale cosmic ray density gradient on the equatorial plane, thus linking the observed anisotropy with possible contributions of the closest sources, such as the Vela supernova remnant at a distance of about 0.3 kpc and with an age of about
11 kyr~\citep{Ahlers:2016rox}. The fact that Vela is located within the large-scale excess region of the sky is consistent with it being a potential source contributing to the large-scale anisotropy. 
However, predictions of the anisotropy amplitude depend on many unknown factors such as the actual contributing source (or sources), the diffusion coefficient, and the unknown component of the anisotropy perpendicular to the equatorial plane that complicate such calculations.

The measured amplitude and phase in this study is consistent with
observations from multiple experiments that show a turning point in the energy dependency of the dipole component amplitude at an energy scale of 10~TeV (see Figure~\ref{fig:phase}). After initially increasing with energy, the dipole amplitude begins to decrease above 10~TeV, while the phase has an abrupt change at the 100~TeV energy scale where the amplitude begins to increase again. Cosmic rays with rigidity of 10~TV have a gyro-radius of about 700~AU in a 3 $\mu$G magnetic field, which is comparable to the transversal size of the heliosphere (i. e. perpendicular to the long axis)~\citep{Pogorelov:2009may,Pogorelov:2013jul,1742-6596-719-1-012013}. It is reasonable to assume that at lower energies the heliospheric influence is important, while above 10~TV the interstellar influence is progressively more important~\citep{desilaza2013}. 
An understanding of how interstellar propagation of 10~TV-scale cosmic rays influences the arrival direction distribution must, therefore, also take into account heliospheric effects~\citep{Schwadron2014,0004-637X-842-1-54, 2016JPhCS.767a2027Z}. An alternative approach is to study cosmic ray anisotropy above 100 TV rigidity~\citep{IceCube:2013mar, icecube2016}, where the heliospheric influence is expected to be negligible. In this case, the arrival direction distribution can be used to probe the global properties of interstellar turbulence by fitting theoretical models to observations~\citep{0004-637X-835-2-258}. However, at high energies a full-sky study is currently not possible with the dataset used in this analysis due to limited statistics.

\subsection{Large-scale Anisotropy and the Local Interstellar Magnetic Field}
\begin{figure}[ht]
\begin{center}
\includegraphics[width=8.5cm]{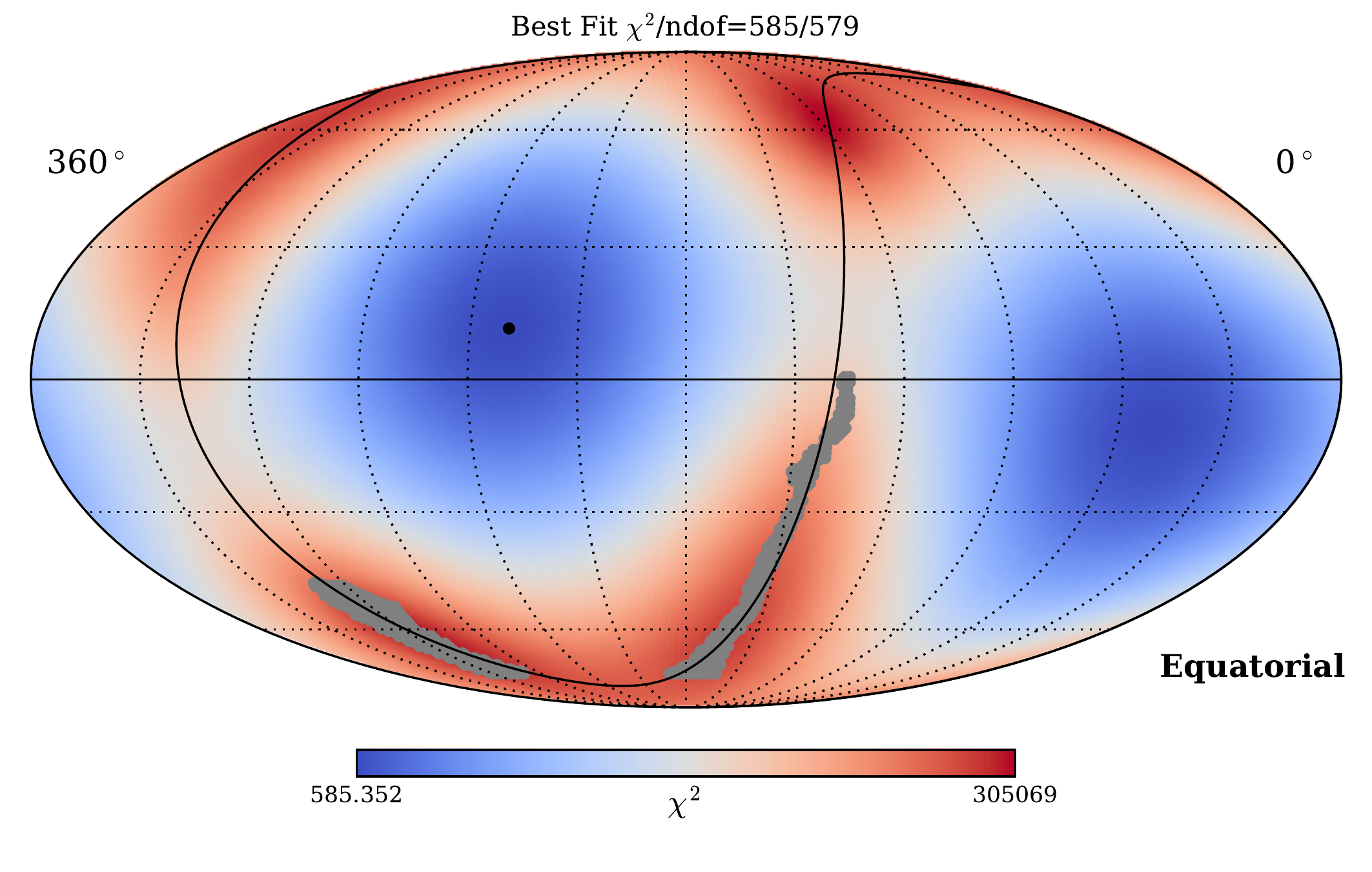}
\caption{$\chi^2$ distribution map for circular fit to boundary between large-scale excess and deficit regions shown in J2000 equatorial coordinates. The black point corresponds to the minimum $\chi^2$ for the center of the circle and the black curve is the fitted circle. The grey points are the selected pixels for the fit. The best fit has a value of $\chi^2$/ndof $=585/579$.      
}
\label{fig:chi2fit}
\end{center}
\end{figure}
\begin{figure}[h]
\begin{center}
\includegraphics[width=0.5\textwidth]{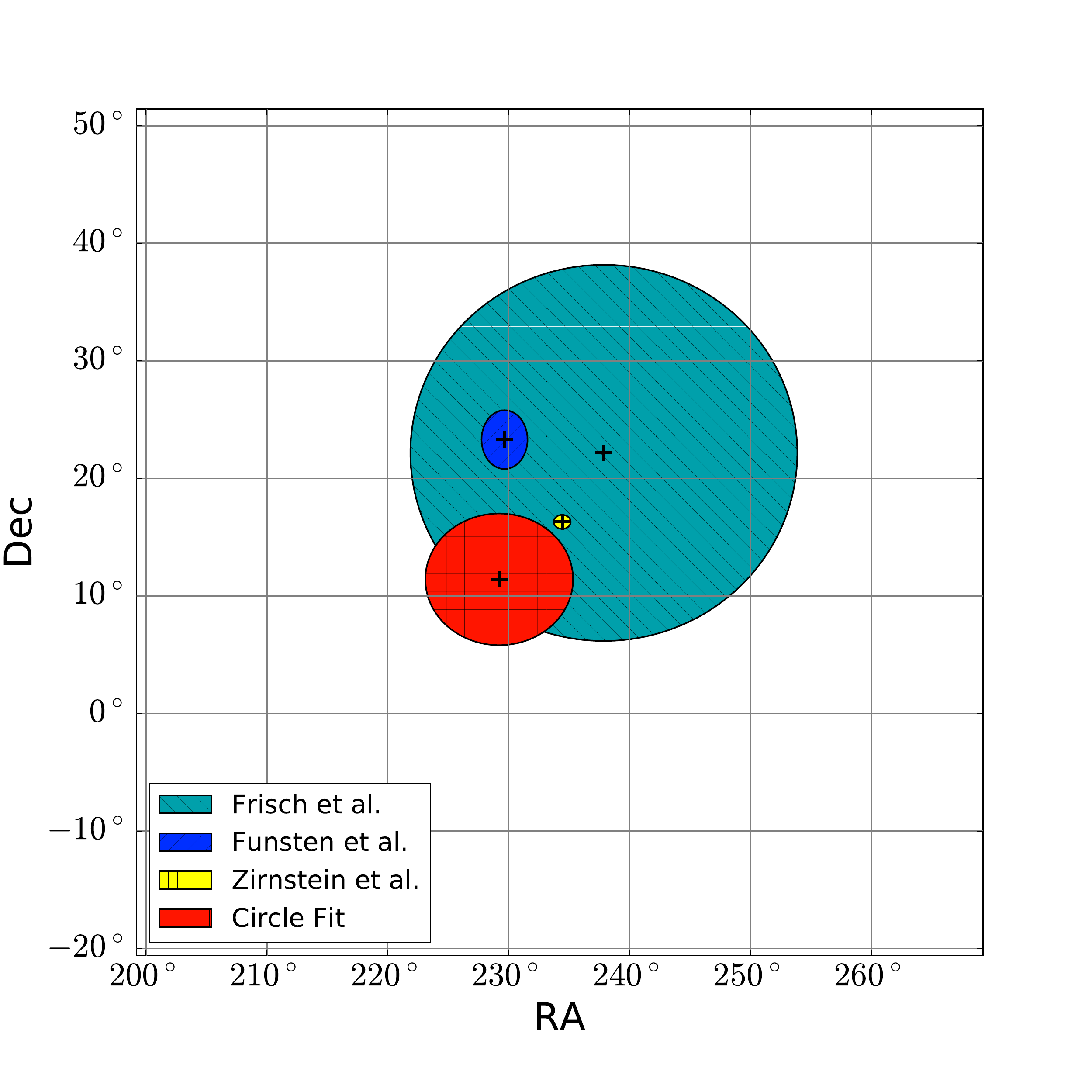}
\caption{Circular fit to boundary between large-scale excess and deficit regions shown in J2000 equatorial coordinates along with published magnetic field measurements by~\cite{0004-637X-776-1-30} 
inferred from the emission of energetic neutral atoms (ENA) originating from the outer heliosphere by the Interstellar Boundary Explorer (IBEX)~\citep{2041-8205-818-1-L18}, and
~\cite{0004-637X-814-2-112} obtained from the polarization of stars within 40 pc. 
}
\label{fig:circlefit}
\end{center}
\end{figure}
Figure~\ref{fig:map:comb_feat} shows the direction of the LIMF from~\cite{2041-8205-818-1-L18} and the corresponding equator (the continuous black line), the so-called $B-V$ plane defined by the LIMF and the direction of the Sun's velocity through the interstellar medium $v_{ISM}$, as well as the direction of the velocity relative to the local standard of rest $v_{LSR}$. The figure also shows the location of the Geminga and Vela supernova remnants as possible contributing sources, and those of the Cygnus X-1 X-ray binary and Galactic center GC for reference. The location of the Galactic plane is shown as a red line.
A fit to the plane defined by the small-scale feature that marks the boundary between the excess and deficit regions ($\sim 115^\circ$ R.A.) is shown in Figure~\ref{fig:chi2fit}. The fit yields a vector pointed towards 
$(\alpha_\mathrm{fit},\delta_\mathrm{fit}) = (229.2\pm 3^\circ.5,  11.4\pm 3^\circ.0)$
in J2000 equatorial coordinates, as shown in Figure~\ref{fig:map:comb_feat}, along with the corresponding equator (the crossed black curve). 
The direction is located $9^\circ$ from the LIMF inferred by the Interstellar Boundary Explorer (IBEX) from the emission of energetic neutral atoms (ENA) originating from the outer heliosphere (see ~\cite{0004-637X-776-1-30}. This point is also located $6^\circ.5$ from the LIMF direction 
reported by~\cite{2041-8205-818-1-L18}
and consistent with the average LIMF direction 
obtained from the polarization of stars within 40 pc by~\cite{0004-637X-814-2-112}. 
This is shown in Figure~\ref{fig:circlefit} and summarized in Table \ref{tab:bismfits} along with the value of $\alpha$ obtained from the dipole fit and the value of $\delta$ obtained from the quadrupole fit. The errors on the fit are derived from the $\chi^2$ distribution shown in Figure~\ref{fig:chi2fit} and don't include possible systematics uncertainties from the missing $m=0$ dipole component.

\begin{table*}[ht]
\centering  
\begin{tabular}{l|l|l|l|l}
Source & R.A. [$^\circ$] & decl. [$^\circ$] & $\Delta \psi$ [$^\circ$] & $\delta_N$ [$10^{-4}$] \\ \hline
\cite{0004-637X-776-1-30} & $229.7\pm 1.9$ & $23.3 \pm 2.5$ & 9.0 & -5.03 \\
\cite{0004-637X-814-2-112} & $237.9\pm16$ & $22.2\pm16$ & 12.2 & -4.77 \\
\cite{2041-8205-818-1-L18} & $234.4 \pm 0.7$ & $16.3\pm 0.6$ & 6.5 & -3.42 \\
Boundary Fit & $229.2\pm 3.5$ & $11.4\pm 3.0$  & -- & -2.36 \\
Dipole/Quadrupole & $218.4 \pm 0.3$ ($\pm 2.6$) & $20.7 \pm 0.3$ ($\pm 2.6$) & -- & -4.41 \\
\end{tabular}
\caption{
\small Magnetic field alignment. The last two rows correspond to measurements of the large-scale anisotropy from this study. The R.A. measurement in the last row is obtained from the dipole vector and the decl. is obtained from the $\ell=2$ quadrupole component. The second to last column corresponds to the angular distance $\Delta \psi$ between the boundary fit and the various LIMF estimates.
The last column gives the corresponding vertical dipole component under the assumption that the dipole is oriented towards the given decl. Error in parentheses for dipole and quadruple correspond to systematic uncertainties.
} \label{tab:bismfits}
\end{table*} 
The fact that the dipole component of the full-sky cosmic ray anisotropy map is approximately aligned with the direction of the LIMF (or at least its projection on the equatorial plane) is probably not a coincidence, since we expect diffusion to be anisotropic with the fastest propagation along the magnetic field lines~\citep{2012A&A...547A.120E, 2014ApJ...785..129K, Schwadron2014, Mertsch:2015jan}.
Assuming that the observed dipole points in this direction, it is possible to estimate the amplitude of the vertical component. The measured amplitude of the horizontal component of the dipole $\tilde{A}_1$ is related to the true amplitude $A_1$ through the dipole inclination $\delta_0$ with $\tilde{A}_1 = A_1 \cos \delta_0$, from which we obtain a value for the vertical dipole vector component of 
$\delta_N = \tilde{A}_1 \tan{\delta_0} \sim -3.97 ^{+1.0}_{-2.0} \times 10^{-4}$ for the various magnetic field assumptions (see Table \ref{tab:bismfits}).

If we assume that the dipole component must be aligned with the LIMF, the observed 
deviation could be explained as due to the relative motion of the observer with respect to a frame in which the cosmic ray distribution is isotropic, called the Compton-Getting Effect~\citep{Compton:1935,1968Ap&SS...2..431G}. 
The heliosphere could also have a significant warping effect on 10 TeV cosmic ray arrival direction distribution, mostly due to the LIMF draping curvature around the heliosphere 
\citep{Pogorelov:2009may}. 
As a result, the dipole component of the cosmic ray anisotropy could be out of alignment from the LIMF. 
Future studies, with full-sky maps at different particle rigidities, could provide a more powerful tool to probe the properties of the interstellar 
and heliospheric magnetic fields.

\section{Conclusions}

We have used experimental data collected by the HAWC Gamma Ray Observatory and the IceCube Neutrino Observatory to compile, for the first time, a nearly full-sky map of the arrival direction distribution of cosmic rays with median energy of 10 TeV. 
The combined analysis accounts for the difference in instantaneous and time-integrated field of view of the HAWC observatory and provides an integrated field of view that extends from $-90^\circ$ to  $+76^\circ$ in decl. The almost full-sky observation eliminates the degeneracy between the spherical harmonic components and provides a tool to probe the properties of particle diffusion in the interstellar medium and of interstellar magnetic turbulence.
The corresponding angular power spectrum suggests that two different mechanisms are responsible for the observed angular scale features. The ordering of cosmic ray anisotropy along the LIMF is supported by fitting the boundary between deficit and excess, which points to the direction $(\alpha_\mathrm{fit},\delta_\mathrm{fit}) = (229.2\pm 3^\circ.5,  11.4\pm 3^\circ.0)$ that is consistent with various observations. We obtained the phase and amplitude of the dipole component projected onto the equatorial plane to be $\tilde{A}_1=(1.17\pm.01) \times 10^{-3}$, 
$\alpha_1=38.4\pm0^\circ.3$. Based on the assumption that the true dipole is aligned along the LIMF, we estimated the missing vertical component to be $\delta_N \sim -3.97 ^{+1.0}_{-2.0} \times 10^{-4}$.

\section*{Acknowledgements}
Dedicated to the memory of our honorable colleague,
and dear friend Stefan Westerhoff.
%
The IceCube collaboration acknowledges the significant contributions to this manuscript from M. Ahlers, P. Desiati., and J.~C.~D\'{i}az-V\'{e}lez. The HAWC Collaboration acknowledges additional contributions from D.~W.~Fiorino.

The HAWC Collaboration acknowledges the support from: the US National Science Foundation (NSF) the US Department of Energy Office of High-Energy Physics; 
the Laboratory Directed Research and Development (LDRD) program of Los Alamos National Laboratory; 
Consejo Nacional de Ciencia y Tecnolog\'{\i}a (CONACyT), M{\'e}xico (grants 271051, 232656, 260378, 179588, 239762, 254964, 271737, 258865, 243290, 132197, 281653)(C{\'a}tedras 873, 1563, 341), Laboratorio Nacional HAWC de rayos gamma; 
L'OREAL Fellowship for Women in Science 2014; 
Red HAWC, M{\'e}xico; 
DGAPA-UNAM (grants IG100317, IN111315, IN111716-3, IA102715, IN109916, IA102917, IN112218); 
VIEP-BUAP; 
PIFI 2012, 2013, PROFOCIE 2014, 2015; 
the University of Wisconsin Alumni Research Foundation; 
the Institute of Geophysics, Planetary Physics, and Signatures at Los Alamos National Laboratory; 
Polish Science Centre grant DEC-2014/13/B/ST9/945, DEC-2017/27/B/ST9/02272; 
Coordinaci{\'o}n de la Investigaci{\'o}n Cient\'{\i}fica de la Universidad Michoacana; Royal Society - Newton Advanced Fellowship 180385. Thanks to Scott Delay, Luciano D\'{\i}az and Eduardo Murrieta for technical support.

The IceCube Collaboration acknowledges the support from:
USA -- U.S. National Science Foundation-Office of Polar Programs,
U.S. National Science Foundation-Physics Division,
Wisconsin Alumni Research Foundation,
Center for High Throughput Computing (CHTC) at the University of Wisconsin-Madison,
Open Science Grid (OSG),
Extreme Science and Engineering Discovery Environment (XSEDE),
U.S. Department of Energy-National Energy Research Scientific Computing Center,
Particle astrophysics research computing center at the University of Maryland,
Institute for Cyber-Enabled Research at Michigan State University,
and Astroparticle physics computational facility at Marquette University;
Belgium -- Funds for Scientific Research (FRS-FNRS and FWO),
FWO Odysseus and Big Science programmes,
and Belgian Federal Science Policy Office (Belspo);
Germany -- Bundesministerium f\"ur Bildung und Forschung (BMBF),
Deutsche Forschungsgemeinschaft (DFG),
Helmholtz Alliance for Astroparticle Physics (HAP),
Initiative and Networking Fund of the Helmholtz Association,
Deutsches Elektronen Synchrotron (DESY),
and High Performance Computing cluster of the RWTH Aachen;
Sweden -- Swedish Research Council,
Swedish Polar Research Secretariat,
Swedish National Infrastructure for Computing (SNIC),
and Knut and Alice Wallenberg Foundation;
Australia -- Australian Research Council;
Canada -- Natural Sciences and Engineering Research Council of Canada,
Calcul Qu\'ebec, Compute Ontario, Canada Foundation for Innovation, WestGrid, and Compute Canada;
Denmark -- Villum Fonden, Danish National Research Foundation (DNRF), Carlsberg Foundation;
New Zealand -- Marsden Fund;
Japan -- Japan Society for Promotion of Science (JSPS)
and Institute for Global Prominent Research (IGPR) of Chiba University;
Korea -- National Research Foundation of Korea (NRF);
Switzerland -- Swiss National Science Foundation (SNSF).

\software{
HEALPix/healpy (version 1.9.1,~\cite{Gorski:2005apr}), 
CORSIKA (version 7.40,~\cite{corsika}), 
ROOT (version 6.04/12,~\cite{citeulike:363715}), 
Matplotlib (version 1.5.0,~\cite{Hunter:2007}), 
Astropy (version 1.1,~\cite{astropy:2013, astropy:2018}), 
SciPy (version 0.16.1, http://www.scipy.org/),
NumPy (version 1.11.1,\cite{Oliphant:2015:GN:2886196}), 
Python programming language (Python Software Foundation, https://www.python.org/),
PolSpice (version 3.0.3, http://www2.iap.fr/users/hivon/software/PolSpice).
}

\appendix

\section{Generalized Maximum likelihood method for multiple observatories with overlapping fields of view}\label{sec:lh}
The method developed by~\cite{0004-637X-823-1-10} assumes that the detector exposure $\mathcal{E}$ per solid angle and sidereal time $t$ accumulated over many sidereal days can be expressed as a product of its angular-integrated exposure $E$ per sidereal time and relative acceptance $A$  (normalized as  $\int d\Omega A(\Omega) = 1$): 
\begin{equation}\label{eq:Exposure}
\mathcal{E}(t,\varphi,\theta) \simeq E(t)A(\varphi,\theta)\,,
\end{equation}
with the assumption that the relative acceptance of the detector does not strongly depend on sidereal time.

For each observatory, the number of cosmic rays expected from 
an angular element $\Delta\Omega_i$ of the local coordinate sphere corresponding to coordinates $(\theta_i,\varphi_i)$ in a sidereal time interval $\Delta t_\tau$ is
\begin{equation}\label{eq:mu}
  \mu_{\tau i} \simeq I_{\tau i}\mathcal{N}_\tau\mathcal{A}_{i}\,,
\end{equation}
where  $\mathcal{N}_\tau\equiv\Delta t_\tau\phi^{\rm iso}{E}(t_\tau)$ gives the expected number of isotropic events in sidereal time bin $\tau$ independent of pixel,  
$\mathcal{A}_i$ is the relative acceptance of the detector for pixel $i$, and $I_{\tau i}\equiv {I}({\bf R}(t_\tau){\bf n}'(\Omega_\mathfrak{i}))$
is the relative intensity observed in local coordinates during time bin $\tau$. ${\bf R}(t){\bf n}' = {\bf n}$ is the time-dependent coordinate transformation of the unit vector ${\bf n}$ that corresponds to the coordinates $(\alpha,\delta)$ in the right-handed equatorial system. 
Here, we adopt the convention used by~\cite{0004-637X-823-1-10} where {\it roman} indices ($i$, $j$) refer to pixels in the local sky map 
and {\it fraktur} indices ($\mathfrak{a}$) refer to pixels in the celestial sky map 
while time bins are indicated by {\it greek} indices ($\tau$, $\kappa$). 
The data observed at a fixed sidereal time bin $\tau$ 
can be described in terms of the observation in local horizontal sky with bin $i$ as $n_{\tau i}$ or transformed into the celestial sky map with bin $\mathfrak{a}$ as $n_{\tau \mathfrak{a}}$.

The likelihood of observing $n$ cosmic rays is then given by the product of Poisson probabilities
\begin{equation}\label{eq:LH}
  \mathcal{L}(n|I,\mathcal{N},\mathcal{A}) =
  \prod_{\tau i}\frac{(\mu_{\tau i})^{n_{\tau i}}e^{-\mu_{\tau i}}}{n_{\tau i}!}\,,
\end{equation}
where $n = \sum_{i, \tau} n_{\tau i}$. 
We maximize the likelihood ratio
of signal over null hypothesis of no anisotropy ($I^{(0)}$, $\mathcal{N}^{(0)}$, $\mathcal{A}^{(0)}$),
\begin{equation}\label{eq:LHratio}
  \lambda = \frac{\mathcal{L}(n|I,\mathcal{N},\mathcal{A})}
                 {\mathcal{L}(n|I^{(0)},\mathcal{N}^{(0)},\mathcal{A}^{(0)})} \,, 
\end{equation}
with $I_\mathfrak{a}^{(0)}=1$. The maximum likelihood estimators of $\mathcal{A}_i$ and $\mathcal{N}_\tau$ are then
\begin{align}\label{eq:Nnull}
  \mathcal{N}_\tau^{(0)} &=  {\sum_i n_{\tau i}}\,,\\\label{eq:Anull}
  {\mathcal{A}}_i^{(0)} &= \sum_\tau n_{\tau i}\Big/\sum_{\kappa j}n_{\kappa j}\,,
\end{align}
given the boundary condition 
\begin{equation}
\sum_i\mathcal{A}_i=1\,.
\end{equation}

In this combined analysis of HAWC and IceCube data, the likelihood (Eq. \ref{eq:LH}) is generalized to a product over data sets with individual detector exposures 
but the {\it same} relative intensity. 
%
The total accumulated exposure $\mathcal{E}$ in Eq.~(\ref{eq:Exposure}) becomes a sum over disjoint sky {\it sectors}, whose union covers the entire field of view. In this analysis the integrated field of view of each detector corresponds to a sector. As before, we assume that the exposure in each sector can be expressed as a product of its angular-integrated exposure $E^s$ and relative acceptance in terms of azimuth $\varphi$ and zenith angle $\theta$ as
\begin{equation}\label{eq:Egen}
  \mathcal{E}(t,\varphi,\theta) \simeq \sum_{{\rm sector}\,{s}}E^{s}(t)\mathcal{A}^{s}(\varphi,\theta)\,.
\end{equation}
The values of $I$, $\mathcal{N}$, and $\mathcal{A}$ of the maximum likelihood ratio (Eq. \ref{eq:LHratio}) $(I^\star,~\mathcal{N}^\star,~\mathcal{A}^\star)$ must obey the implicit relations

\begin{align}\label{eq:Istar}
  {I}^\star_{\mathfrak{a}} &=
  \sum_{\tau} n_{\tau\mathfrak{a}}\Big/ \sum_{{s}\kappa}\mathcal{A}^{s\,\star}_{\kappa \mathfrak{a}}\mathcal{N}^{{s}\,\star}_\kappa \,,\\
  \label{eq:Nstar}
  \mathcal{N}^{{s}\,\star}_\tau &=
  \sum_{i} w^{s}_i n_{\tau i}\Big/\sum_{j}\mathcal{A}^{s\,\star}_jI^\star_{\tau j}\,,\\
  \label{eq:Astar}
\mathcal{A}^{s\,\star}_i &= \sum_\tau w^{s}_i n_{\tau i}\Big/\sum_{\kappa}\mathcal{N}^{s\,\star}_\kappa I^\star_{\kappa i}\,.
\end{align}
Here, we have introduced the window function $w^s_i$ of the sector $s$ which is equal to $1$ if the pixel $i$ is located in the sector and $0$ otherwise.
The binned quantity $\mathcal{A}^s_{\tau\mathfrak{a}}$ 
in Eq.~(\ref{eq:Istar}), corresponds to the relative acceptance of sector $s$ seen in the equatorial coordinate system in pixel $\mathfrak{a}$ during time bin $\tau$.
Equations~\eqref{eq:Istar} to \eqref{eq:Astar} correspond to a nonlinear set of equations that cannot be solved in explicit form 
but one can iteratively approach the best-fit solution.

This reconstruction method is a simple generalization of the iterative method outlined in
~\cite{0004-637X-823-1-10}, where now the relative acceptances $\mathcal{A}$ and isotropic expectation $\mathcal{N}$ for each detector are evaluated as independent quantities.  This is a valid approach as long as the rigidity distributions of the data sets are very similar. 

\bibliography{IC86_HAWC_CR_Aniso}
\bibliographystyle{apj}



\end{document}